\documentclass[%
 reprint,
superscriptaddress,
nofootinbib,
 amsmath,amssymb,
 aps,
 prb,
]{revtex4-2}

\usepackage{graphicx}
\usepackage{dcolumn}
\usepackage{bm}

\usepackage[protrusion=true,expansion=true]{microtype}
\usepackage{times}

\usepackage{subcaption}
\usepackage{ragged2e}
\DeclareCaptionJustification{justified}{\justifying}
\usepackage[utf8]{inputenc}
\usepackage{geometry}
\usepackage[dvipsnames]{xcolor}
\usepackage{booktabs}
\usepackage{slashed}
\usepackage{multirow}

\usepackage{mathtools}
\def\block(#1,#2)#3{\multicolumn{#2}{c}{\multirow{#1}{*}{$ #3 $}}}

\usepackage{amsmath}
\usepackage{amsthm}
\usepackage{amssymb}
\usepackage{float}
\usepackage{graphicx}
\usepackage[british]{babel}
\usepackage{braket}
\usepackage{textcase}
\usepackage{setspace}
\usepackage{enumerate}
\usepackage[normalem]{ulem}
\usepackage[final]{hyperref} 
\hypersetup{
	colorlinks=true,       
	linkcolor=red,        
	citecolor=red,        
	filecolor=magenta,     
	urlcolor=blue         
}
\usepackage{verbatim}



\begin{document}
\preprint{}

\title{\textbf{Charge-resolved entanglement in the presence of topological
defects}}
\author{D\'avid X. Horv\'ath}

\affiliation{SISSA and INFN Sezione di Trieste, via Bonomea 265, 34136 Trieste, Italy}
\author{Shachar Fraenkel}
\affiliation{Raymond and Beverly Sackler School of Physics and Astronomy, Tel Aviv
University, Tel Aviv 6997801, Israel}
\author{Stefano Scopa}

\affiliation{SISSA and INFN Sezione di Trieste, via Bonomea 265, 34136 Trieste, Italy}
\author{Colin Rylands}

\affiliation{SISSA and INFN Sezione di Trieste, via Bonomea 265, 34136 Trieste, Italy}

\begin{abstract}
Topological excitations or defects such as solitons are ubiquitous throughout physics, supporting numerous interesting phenomena like zero energy modes with exotic statistics and fractionalized charges.  In this paper, we study such objects through the lens of symmetry-resolved entanglement entropy.  Specifically, we compute the charge-resolved entanglement entropy for a single interval in the low-lying states of the Su-Schrieffer-Heeger model in the presence of topological defects. Using a combination of exact and asymptotic analytical techniques, backed up by numerical analysis, we find that, compared to the unresolved counterpart and to the pure system, a richer structure of entanglement  emerges. This includes a  redistribution between its configurational and fluctuational parts due to the presence of the defect and an interesting interplay with entanglement equipartition. 
In particular, in a subsystem that excludes the defect, equipartition is restricted to charge sectors of the same parity, while full equipartition is restored if the subsystem includes the defect, as long as the associated zero mode remains unoccupied. Additionally,
by exciting zero modes in the presence of multiple defects, we observe a significant enhancement of entanglement in certain charge sectors,
due to charge splitting on the defects. The two different scenarios featuring the breakdown of entanglement equipartition are underlied by a joint mechanism, which we unveil by relating them to degeneracies in the spectrum of the entanglement Hamiltonian. In addition, equipartition is shown to stem from an equidistant entanglement spectrum.
\end{abstract}

\maketitle


\section{Introduction}

Interfaces between different phases of matter are a bountiful hunting ground for interesting physical phenomena.  These can occur in parameter space,  wherein by tuning a control parameter such as temperature one can enter a critical region exhibiting universal properties and gapless excitations, or in real space by joining systems with distinct properties.  An especially intriguing example of the latter is when the systems are in different topological phases.  In these cases the interfaces can also host gapless excitations which endow distinct universal properties on the system. For example, the edges of topological insulators host metallic states resulting in quantized transport properties~\cite{qi2011topological}.  This behvaior is not limited to boundaries and can also occur in the bulk through the introduction of certain topological defects.  

The archetypal example of this is given in the Su-Schreiffer-Heeger (SSH) model~\cite{su1979solitons} (see Ref.~\cite{Asb_th_2016} for a pedagogical review) or its field theoretic counterpart, the Jackiw-Rebbi (JR) model~\cite{jackiw1976solitons}.  The SSH model describes free fermions on a lattice with the intersite hopping alternating between weak and strong, while the JR model can be viewed as the low-energy, continuum limit of the SSH describing a massive Dirac fermion~\cite{jackiw1981solitons}.  A topological defect known as a soliton can be introduced in the former by changing the pattern of the hopping strengths from e.g.~weak-strong to strong-weak at a certain lattice site, and in the latter by tuning the mass of the fermion so that it changes sign at a certain point.  These solitons mark a transition between a topological and a non-topological phase and accordingly have a number of notable features.  In particular, they host localized zero-energy modes, which are characteristic of a transition between topological and non-topological phases, and they also carry fractional charges~\cite{wilczek1981fractional}.  In the continuum limit, these fractional charges can be understood as a fractionalization of the charge symmetry due to the presence of the soliton, so that the local charge operator carries fractional eigenvalues rather than just fractional expectation values~\cite{jackiw1983fluctuations}. The zero modes are zero-energy excitations above this fractionally charged background which themselves do not carry fractional charge.
They are related to Majorana zero modes, which have been the focus of much attention in the last two decades from a fundamental physics perspective and for their potential application to fault tolerant quantum computing~\cite{kitaev2001unpaired,alicea2012new}.  

A key diagnostic tool used in studies of topological systems is the entanglement entropy between a subsystem and its complement~\cite{li2008entanglement}.  By examining the spectrum of the reduced density matrix of this subsystem, one can determine whether the system is in a topological or trivial phase.  Aside from this, the entanglement entropy provides key information on numerous properties of a system both in and out of equilibrium.  Recently, the introduction of symmetry-resolved (either R\'enyi or von Neumann) entanglement entropies (SREs) \cite{PhysRevLett.120.200602,Laflorencie_2014} allowed to better investigate the interplay between entanglement and symmetries, becoming the topic of several studies. 

SREs allow to study how entanglement varies between different symmetry sectors and by now have been studied in many different scenarios leading to some notable results, such as the equipartition of entanglement entropy and a time delay for the spreading of entanglement after a quantum quench~\cite{PhysRevLett.120.200602,Laflorencie_2014,xavier2018equipartition,PhysRevA.98.032302, parez2021quasiparticle,LAFLORENCIE20161, bonsignori2019symmetry, murciano2020entanglement,Murciano_2020,CalabreseFCS_2020,PhysRevB.102.014455,Horv_th_2020, Horv_th_2021, Capizzi_2022,Castro_Alvaredo_2023,MicheleNonHermitian,Monkman_2020,Cornfeld_2019, Azses_2020,Chen_2021,digiulio2023boundary,Calabrese_2021,murciano2023symmetryresolved,Horvath_2022,Ares_2022_MultiCharge,Ares_2022,Feldman_2019,parez2021exact,fraenkel2020symmetry,Scopa_2022,PhysRevB.101.235169,10.21468/SciPostPhys.10.3.054,bertini2023nonequilibrium,bertini2023dynamics, Ghasemi_2023, capizzi2023counting,foligno2023entanglement,murciano2022page,murciano2021fermion,digiulio2023symmetry,PhysRevB.107.115113,PhysRevB.107.014308}.
The study of entanglement across defects in low-dimensional systems has also become a topic of interest lately, with several studies exploring how the presence of a defect can affect the equilibrium and non-equilibrium properties of a system and how this is manifest in the entanglement entropy \cite{PhysRevLett.128.090603,fraenkel2021entanglement,Alba_SciPost_2022, fraenkel2022extensive,Caceffo_2023,rylands2023transport, capizzi2023domain, capizzi2023entanglement, capizzi2023zeromode}.
 
In this work, we initiate a study at the intersection of these topics and examine symmetry-resolved entanglement entropy across topological defects. In particular, using a combination of exact and asymptotic analytical techniques complemented by numerical analysis, we study the charge-resolved entanglement in the SSH  model containing solitons. We compare the cases where the subsystem of interest is located in the topological or trivial phases, and the case where it straddles the defect. We find that the unresolved quantities display a rather mundane picture for such defects~--~the total entropy seemingly being obtained by counting the number of strong bonds that are cut by the subsystem boundaries~\cite{ryu2006entanglement,micallo2020topological}.  In contrast, the charge-resolved quantities present a much richer structure when the subsystem contains a soliton, with entanglement redistributed between its configuration and fluctuation parts.

The soliton also exhibits an interesting interplay with the equipartition of entanglement. Entanglement equipartition, i.e., the phenomenon where SREs are independent (to a leading order in subsystem size) of the charge sector -- is known to occur with respect to a $U\!\left(1\right)$ symmetry in critical ground states \cite{xavier2018equipartition,PhysRevB.102.014455,bonsignori2019symmetry,fraenkel2020symmetry} and in ground states of massive field theories \cite{murciano2020entanglement,Horv_th_2021,Horvath_2022}, but has been observed to break down in certain gapped ground states of lattice systems \cite{Murciano_2020,CalabreseFCS_2020}.

For the ground state we consider here, we find that entanglement equipartition is broken in a pure subsystem, yet it can in fact be restored by the presence of the defect. We also explore the effect of zero modes on SREs in the presence of multiple defects. Upon charge resolution, we observe a significant enhancement of the entanglement entropy in certain charge sectors due to degeneracies and level crossings of the entanglement Hamiltonian. Entanglement equipartition is therefore absent in this case also. As the precise conditions for equipartition remain unclear, especially for massive theories, we aim to shed light on this question by studying a model where equipartition and its breakdown cohabit.

Our results highlight a hidden entanglement structure that can arise in the presence of topological defects which is only revealed upon symmetry resolution.  Although we focus on the SSH model, we expect similar results to hold also for the JR model. Moreover, even though the SSH and JR models are non-interacting and contain only static solitons, they might be viewed as self-consistent mean field solutions of interacting systems containing solitons as dynamical excitations.  Therefore, our results can provide useful insight for their analysis, as argued below. \\

The remainder of the paper is structured as follows. In Sec.~\ref{sec:2} we briefly review the notion of symmetry-resolved entanglement entropy and set up some basic notation.  Following this, in Sec.~\ref{sec:3} we introduce the Hamiltonian and discuss its properties.  In Sec.~\ref{sec:4} we establish some basic intuition for our results by examining the fully dimerized limit of the model, after which in Sec.~\ref{AwayFromDimerized} we perform an analysis away from it using a combination of asymptotically exact analytical results backed up by numerics.  In Sec.~\ref{SectionWithZM}, we examine the influence of the zero modes on the entanglement spectrum, and in particular
the way in which the results are modified if the zero modes change from being localized on a single defect to a two-defect localized mode, that is, when the zero mode has (not necessarily equal) support on two defects. Finally, in Sec.~\ref{sec:conclusions} we comment on the applicability of these results to other models, including interacting ones, and conclude. 

\section{Symmetry Resolved Entanglement}\label{sec:2}
As is  widely known, when a system is in a pure state, the bipartite entanglement
of a subsystem $A$ and its complement $\bar{A}$ can be quantified by the R\'enyi entanglement entropies
\cite{RevModPhys.80.517,Calabrese_2009,RevModPhys.82.277,LAFLORENCIE20161} 
\begin{equation}
S_{n}=\frac{1}{1-n}\log\text{Tr}\rho_{A}^{n}=\frac{1}{1-n}\log\mathcal{Z}_n\,,
\label{TotalRenyi}
\end{equation}
defined in terms of the reduced density matrix (RDM) $\rho_{A}$ or the partition function $\mathcal{Z}_n=\text{Tr}\rho_{A}^{n}$ of
the subsystem $A$.  From those, in the replica limit $n\rightarrow1$
the von Neumann entropy 
\begin{equation}
S=-\text{Tr}\rho_{A}\log\rho_{A}
\label{TotalVN}
\end{equation}
is obtained. 

The  idea of explicitly considering the internal
structure of entanglement associated with symmetry was introduced in Refs.~\cite{Laflorencie_2014,PhysRevLett.120.200602}. In a symmetric
state, the conserved charge $\hat{Q}$ corresponding to the symmetry commutes with the density matrix. Under general circumstances, we may decompose $\hat{Q}=\hat{Q}_A\oplus \hat{Q}_{\bar{A}}$, with $\hat{Q}_B$ acting only in region $B=A,\bar{A}$. From this, one finds 
 that  $\hat{Q}_A$ commutes with the RDM, 
\begin{equation}
[\rho_{A},\hat{Q}_{A}]=0\,.\label{eq:Commutation}
\end{equation}
Such commutation implies that $\rho_{A}$ is block-diagonal, each
block corresponding to an eigenvalue of $\hat{Q}_{A}$. Consequently, the R\'enyi and von Neumann entropies
can be decomposed according to the symmetry sectors of $\hat{Q}_{A}$.
The symmetry-resolved R\'enyi and von Neumann entropies can be eventually defined
as 
\begin{equation}
S_{n}(q_{A})=\frac{1}{1-n}\log\!\left[\frac{\mathcal{Z}_{n}(q_{A})}{\mathcal{Z}_{1}^{n}(q_{A})}\right]
\label{eq:RE}
\end{equation}
and
\begin{equation}
S(q_{A})=-\frac{\partial}{\partial n}\left[\frac{\mathcal{Z}_{n}(q_{A})}{\mathcal{Z}_{1}^{n}(q_{A})}\right]_{n=1}\label{eq:vNE}
\end{equation}
in terms of the symmetry-resolved partition functions (SRPFs)
\begin{equation}
\mathcal{Z}_{n}(q_{A})=\text{Tr}\left(\rho_{A}^{n}\mathcal{P}(q_{A})\right)\,,
\label{SRPF_Definition}
\end{equation}
where $\mathcal{P}(q_{A})$ is the projector onto the sector corresponding
to the eigenvalue $q_{A}$ of $\hat{Q}_A$.  Herein $\mathcal{Z}_{1}(q_{A})\leq 1$ is the probability that a measurement of $\hat{Q}_A$ returns the value $q_A$.  It appears in \eqref{eq:RE} and~\eqref{eq:vNE} so as to properly normalize the reduced density matrix in each charge sector. 

The calculation of these symmetry-resolved quantities would require,
in general, the diagonalization of $\rho_{A}$ and the resolution of
the spectrum in the conserved charge. However, an ingenious way to circumvent this difficulty is the introduction of the charged moments \cite{PhysRevLett.120.200602},
defined for a $U(1)$ symmetry as 
\begin{equation}
Z_{n}(\alpha)=\text{Tr}\left(\rho_{A}^{n}e^{i\alpha\hat{Q}_{A}}\right).\label{eq:Zn}
\end{equation}
These are nothing but the Fourier transform of the desired partition
functions, i.e.~\cite{PhysRevLett.120.200602} 
\begin{equation}
\mathcal{Z}_{n}(q_{A})=\int_{-\pi}^{\pi}\frac{\mathrm{d}\alpha}{2\text{\ensuremath{\pi}}}Z_{n}(\text{\ensuremath{\alpha}})e^{-i\alpha q_{A}}.\label{eq:CalU1}
\end{equation}
The charged moments can be readily determined using well-established methods on R\'enyi entropies, and their Fourier transform can be computed thereafter.

A notable feature of SREs is that, in the case of the von Neumann entanglement measures, we can relate the symmetry-resolved and the total (unresolved) quantities as
\begin{equation}
S=\sum_{q_{A}}\mathcal{Z}_1(q_{A})S(q_{A})-\sum_{q_{A}}\mathcal{Z}_{1}(q_{A})\log \mathcal{Z}_{1}(q_{A})=S_{c}+S_{f}\,.\label{eq:SfSc}
\end{equation}
The contribution $S_{c}$ is called the configuration entanglement entropy and amounts to the entropy due to each charge sector weighted with the corresponding probability \cite{Lukin_2019,Wiseman_2003}, and $S_{f}$ denotes the fluctuation (or number) entanglement entropy, which instead takes into account the entropy due to the fluctuations of the value of the charge in the subsystem $A$ \cite{Lukin_2019,Kiefer_Emmanouilidis_2020, Kiefer_Emmanouilidis_2020b}. Symmetry-resolved entanglement recently became accessible in different experimental platforms, see e.g.~Refs.~\cite{PhysRevLett.125.120502,Neven_2021,Vitale_2022,Rath_2023}. 

\section{Paradigmatic model of topological phases: the SSH chain}\label{sec:3}
\begin{figure}[t]
\centering
\includegraphics[width=\columnwidth]{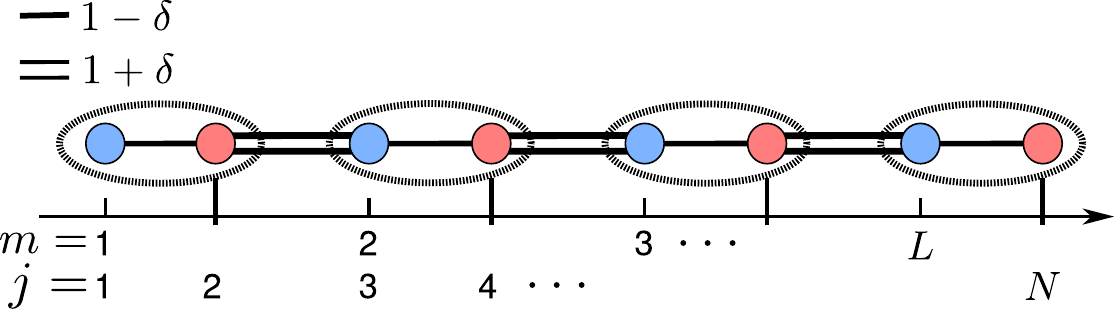}
\caption{\label{FigSSH}Illustration of the SSH model: single and double lines denote hopping amplitudes between neighboring sites of magnitude $1-\delta$ and $1+\delta$, respectively ($t=1$).  
 Lattice sites are indexed by $j=1,\dots, N$ ($N$ is the system size), while we use the index $m=1,\dots,L\equiv N/2$ to refer to "cells" comprising two consecutive sites. The particular alignment of the cells corresponds to the topological phase of the model for $\delta>0$.}
\end{figure}

The SSH model describes the hopping of spinless free fermions on a lattice with an alternating pattern of hopping strength. In terms of the  $c_j^\dag, c_j$  spinless fermion creation and annihilation operators at a lattice site $j$, the Hamiltonian of the SSH model is defined as follows
\begin{equation}
H=-t\sum_{i=1}^{L}\left[\left(1-\delta\right)c_{2i-1}^\dag c_{2i}+\left(1+\delta\right)c_{2i}^\dag c_{2i+1}\right]+\text{H.c.}
\label{SSHHamPBC}
\end{equation}
Here, the hopping strength alternates between being $t\left(1-\delta\right)$ and  $t\left(1+\delta\right)$ (for $\delta>0$ these are, respectively, a weak bond and a strong bond, and vice versa for $\delta<0$).  We call $\delta$ the dimerization parameter, and refer to $\delta=\pm1$ as the fully dimerized case. We set periodic boundary conditions (PBC) on a finite system via $2L+1\equiv1$.
Equivalently, one can also define a three-site Hamiltonian density $h_i\!\left(t,\delta\right)$, written as
\begin{equation}
h_i\!\left(t,\delta\right)=-t\left[\left(1-\delta\right)c_{2i-1}^\dag c_{2i}+\left(1+\delta\right)c_{2i}^\dag c_{2i+1}\right]+\text{H.c.}\,,
\end{equation}
in terms of which the Hamiltonian \eqref{SSHHamPBC} reads as
\begin{equation}
H=\sum_{i=1}^{L} h_i\!\left(t,\delta\right)\,.
\end{equation}
The  peculiar dimerized structure of the Hamiltonian \eqref{SSHHamPBC} is illustrated in Fig.~\ref{FigSSH}.
For the SSH chain, it is customary to introduce the notion of a {\it super-lattice} in the form of unit cells encompassing two neighboring sites. Throughout the paper, we shall label the cell index with $m=1,\dots, L$, and denote the total number of cells with $L$. For (elementary) lattice sites, we shall instead use the letters "$i$" and "$j$", with $N\equiv 2L$ denoting the system size. Using the super-lattice structure, one can easily and naturally distinguish between the trivial and the topological phases of the system, that is, whether the intra-cell hopping or the inter-cell hopping is stronger, respectively. These correspond to a negative or positive value of the dimerization parameter $\delta$. 
 
The spectrum of the model, when PBC are imposed, is given in terms of the following dispersion relation, consisting of two bands:
\begin{equation}
\varepsilon_{k}=\pm 2t\sqrt{\cos^{2}\!\left(\frac{k}{2}\right)+\delta^{2}\sin^{2}\!\left(\frac{k}{2}\right)}\,.
\end{equation}
We can label the single-particle energies in terms of the momentum
$k$, and the corresponding single-particle states are delocalized plane-wave eigenfunctions. Here, we note that the system has a band gap of $\Delta=4t|\delta|$ at $k=\pm \pi$ and the spectrum is symmetric around $\varepsilon=0$, due to the chiral symmetry of the model. Notice that a well-known fingerprint of topological phases is that a system with finite length hosts zero-energy edge states.
Accordingly, in the topological phase and when open boundary conditions (OBC) are considered, the above-mentioned zero modes are present in the spectrum and they are localized at the edges of the system. The localization length of these modes can be computed from the condition $\varepsilon\!\left(k^*\right)=0$, which gives $k^*=\pm\,2i\, \text{artanh}\!\left(\delta\right)\pm\pi$. When $k^*$ is substituted into the plane-wave function, the localization length is eventually given by 
\begin{equation}
\xi^{-1}= 2\, \text{artanh}\!\left(\delta\right)\,,
\label{LocLength}
\end{equation}
and the envelope of the wave function approximately reads as $\exp\!\left(-m/\xi\right)$. These states are mapped to themselves under the chiral symmetry.\\

A defect in the SSH Hamiltonian can be realized by changing the dimerization pattern starting from a certain site.  At the defect, the phase changes from topological to trivial, and the model hosts a zero-energy mode akin to the edge mode described above.  The localization length of such modes is also given by $\xi$. However, because of the super-lattice structure imposed on the model, when using PBC, only an even number of defects can be present in the system.  

\begin{figure}[t]
\centering
(a)\ \includegraphics[width=0.9\columnwidth]{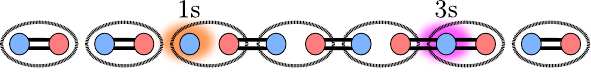}\\
\vspace{0.5cm}
(b)\ \includegraphics[width=0.9\columnwidth]{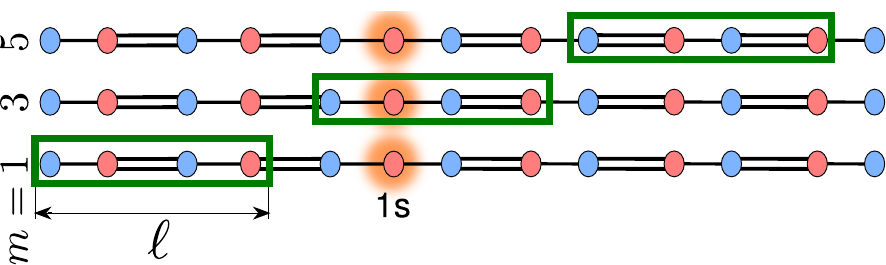}
\caption{\label{figDefTypes} {\it Panel} (a)~--~The cell structure superimposed on the SSH chain. 
The example shown features a one-site (1s) defect and a three-site (3s) defect.
{\it Panel} (b)~--~The setup for the study of SREs in the SSH chain. An interval $[m,m+\ell-1]$ consisting of $\ell$ cells and an initial cell index $m$ making up the subsystem $A$ is moved along the chain. The interval is represented by a green rectangle with length $\ell=2$ and its initial index is $m=1,3$ and $5$. It can be in the trivial ($m=5$) or in the topological phase ($m=1$), or it can contain not more than one defect ($m=3$).}
\end{figure}

For concreteness, we assume without loss of generality that $\delta>0$, and distinguish between one-site (1s) and three-site (3s) defects, although both choices will lead to the same physical results.
A three-site defect (or trimer) is obtained by changing the dimerization pattern as
\begin{equation}
H_{3s}^{\text{def}}= \sum_{i\le j}h_i(t,\delta)+\sum_{i>j} h_i(t,-\delta)\,,
\end{equation}
such that our Hamiltonian is made of three consecutive sites connected by strong bonds. Its name is understood from the fully dimerized limit, for which a 3s defect corresponds to a chain of three connected sites. Note that in the above equation we only specified a segment of the Hamiltonian and hence the lower and upper limits for the summation are not manifest. The other type of defect is
\begin{equation}\label{eq:1s-def}
\begin{split}H_{1s}^{\text{def}}= & \sum_{i\le j-1}h_i(t,\delta)-\left[t(1-\delta)c_{2j-1}^\dag c_{2j}+\text{H.c.}\right]\\
&-\left[t(1-\delta)c_{2j}^\dag c_{2j+1}+\text{H.c.}\right]+\sum_{i\geq j+1}h_i(t,-\delta)\,,
\end{split}
\end{equation}
consisting of three consecutive sites connected by weak bonds. We refer to it as a one-site (1s) defect, given that the corresponding configuration in the fully dimerized limit contains a single site. Similarly, the lower and upper limits for the summation in Eq.~\eqref{eq:1s-def} are not specified. An illustration of the two types of defects for the SSH chain is shown in Fig.~\ref{figDefTypes}(a). 
Regardless of the type of defect considered, each one hosts exactly one zero-energy mode. In addition, the trimer introduces two other modes whose energies are above and below the two bands, forming a particle-hole pair. 

\section{Symmetry-resolved entropies for the SSH chain: the fully dimerized limit}\label{sec:4}

Since the SSH chain describes a non-interacting system of spinless fermions, the computation of the reduced density matrices for states in the form of a Slater determinant can be obtained by considering the correlation matrix $C_{ij}$~\cite{Peschel_2003}, defined as
\begin{equation}
C_{ij}=\langle\Psi| c_i^{\dagger}c_j|\Psi\rangle\,,
\end{equation}
where the indices $i,j\in A$ refer to (elementary) lattice sites in the subsystem, and $|\Psi\rangle$ is the many-body wave function of the system. Due to the Slater determinant form, the above formula can be rewritten in terms of the wave functions $\phi_k$ of the single-particle states that are occupied within the many-body ground state,
\begin{equation}
C_{ij}=\sum_k\langle\phi_k | i\rangle\langle j|\phi_k\rangle\,,
\label{CijSumk}
\end{equation}
where $|i\rangle=c_i^\dag |0\rangle$. Crucially, the single-particle eigenvalues $\epsilon_i$ of the entanglement Hamiltonian $\hat{H}_A=-\log \rho_A$ and the eigenvalues $\lambda_i$ of the correlation matrix are related by Peschel's formula \cite{Peschel_2003}
\begin{equation}
\lambda_i=\frac{1}{\exp\!\left(\epsilon_i\right)+1}\,,
\label{PeschelFormula}
\end{equation}
and the conventional $\mathcal{Z}_{n}$ partition function can be written in terms of the eigenvalues of the correlation matrix as
\begin{equation}
\mathcal{Z}_{n}=\prod_{i}\left(\lambda_i^{n}+\left(1-\lambda_i\right)^{n}\right)\,.
\label{Z_PartitionFunctionCf}
\end{equation}
Regarding the symmetry-resolved quantities, a similar expression can be written for the charged moments defined in Eq.~\eqref{eq:Zn} \cite{PhysRevLett.120.200602}, namely
\begin{equation}
Z_{n}\!\left(\alpha\right)=\prod_{i}\left(\lambda_i^{n}e^{i\alpha}+\left(1-\lambda_i\right)^{n}\right)\,.
\label{Z_ChargedMomentsCf}
\end{equation}
The number of terms in the product equals the length of the subsystem size if the cell structure is not considered, otherwise it is twice the length of the subsystem w.r.t.~the cells. We recall that the SRPFs $\mathcal{Z}_n\!\left(q_A\right)$, defined in Eq.~\eqref{SRPF_Definition}, are obtained from the charged moments via a Fourier transform (see Eq.~\eqref{eq:CalU1}).

In the following, we consider the total and symmetry-resolved entropies for the single interval 
\begin{equation}\label{eq:interval-convention}
A=[m,m+\ell-1]\,,
\end{equation}
which begins at the $m$th unit cell and contains $\ell$ cells.
For transparency, we consider systems with only two defects, and assume throughout the paper that the subsystem $A$ never contains more than one defect. We illustrate this in Fig.~\ref{figDefTypes}(b). \\

In the fully dimerized limit of the SSH chain ($\delta=1$), the expressions for the SREs greatly simplify. 
Although the calculation at $\delta=1$ is somewhat straightforward, we find it instructive as it highlights some of the key insights that were used to tackle the problem when $\delta\neq 1$. This holds true also in the presence of defects as long as the system is at half filling or slightly below it, i.e., when either only one or none of the zero-energy modes are excited. The discussion of the model in the presence of excited zero modes is, however, postponed to Sec.~\ref{SectionWithZM}.

\subsection{Correlation matrices}

In the fully dimerized limit, the properties of the correlation matrix do not depend on $m$, i.e., on the particular starting point of the interval \eqref{eq:interval-convention}, but only on its length $\ell$ and on whether $A$ cuts a portion of the trivial phase (triv), of the topological phase (top), or a change between the two by straddling a defect (def). Irrespective of which of these cases is considered, the correlation matrix is a $2\ell\times2\ell$ matrix and can be simply written in terms of block matrices. We distinguish these cases by denoting the corresponding matrices as $C_\text{triv}$, $C_\text{top}$ and $C_\text{def}$, whose explicit expressions can be found in Appendix~\ref{AppendixCorrelMatrices}.
Given the results for the correlation matrices, it is easy to determine their eigenvalues and hence various entanglement measures. The eigenvalues (with the associated degeneracy) are
\begin{equation}
\begin{split} 
& \text{\text{spect}}(C_\text{triv})=
\begin{cases}
0 & \,\,\,\,\,\, \text{deg}:\ell\\
1 & \,\,\,\,\,\, \text{deg}:\ell
\end{cases}\\
& \text{\text{spect}}(C_\text{top})=
\begin{cases}
0 & \text{deg}:\ell-1\\
1/2 & \text{deg}:2\\
1 & \text{deg}:\ell-1
\end{cases}\\
&\text{\text{spect}}(C_{\text{def}})=
\begin{cases}
0 & \text{deg}:\ell\\
1/2 & \text{deg}:1\\
1 & \text{deg}:\ell-1\,.
\end{cases}
\end{split}
\end{equation}
We can see that the two types of defect give rise to the same multiplicities (see Appendix~\ref{AppendixCorrelMatrices}).

\subsection{Total entropies}
\label{FullyDimerizedTotalEntropies}

Given the eigenvalues of the correlation matrices, the partition functions $\mathcal{Z}_{n}$ for the 
interval $A$ can be
written as
\begin{equation}
\mathcal{Z}_{n}=\begin{cases}
0 & \text{triv}\\
\left(\frac{1}{2}\right)^{2n-2} & \text{top}\\
\left(\frac{1}{2}\right)^{n-1} & \text{def}\,.
\end{cases}
\end{equation}
From here we easily obtain 
\begin{equation}
S_{n}=\begin{cases}
0 & \text{triv}\\
2\log2 & \text{top}\\
\log2 & \text{def}\,,
\end{cases}
\end{equation}
that is, the von Neumann entropy and all the R\'enyi entropies exhibit the same behavior. This result is very natural, as is its physical interpretation: the entropy of an interval can be obtained by counting cut (strong) bonds, and each such bond contributes with a $\log2$. In the topological phase, two bonds are cut, while in the trivial phase no bond is cut.
When the defect is contained in the interval, the dimerization pattern changes between the two boundaries of the interval, and therefore only one bond is cut. 

Therefore, the total entropy of an interval with a defect is essentially the average of the corresponding entropies in the trivial and topological phases. From the point of view of the total entropies, the defect is thus a rather mundane object. This finding shall be demonstrated to hold also away from the fully dimerized limit in the subsequent section. Despite this, as we shall shortly discuss, the symmetry-resolved versions of the entropies display a refined (and more illuminating) structure depending on the location of the subsystem.

\subsection{Symmetry-resolved entropies}\label{SREFullyDimerised}
We start our analysis by computing the charged moments, which are
\begin{equation}
Z_{n}(\alpha)=\begin{cases}
e^{i\alpha\ell} & \text{triv}\\
e^{i\alpha\ell}\left(\frac{1}{2^{2n}}e^{i\alpha}+\frac{1}{2^{2n-1}}+\frac{1}{2^{2n}}e^{-i\alpha}\right) & \text{top}\\
e^{i\alpha(\ell-1/2)}\frac{1}{2^{2n}}\left(e^{i\alpha/2}+e^{-i\alpha/2}\right) & \text{def}\,.
\end{cases}
\label{FullyDimerizedChargedMoments}
\end{equation}
From these, the SRPFs are obtained as follows. First, recall that the average charge $\langle \hat{Q}_A \rangle$ in the subsystem is essentially $\ell$ because the system is (slightly below) half filling. More precisely,
\begin{equation}
\langle \hat{Q}_A \rangle=\begin{cases}
\ell & \text{top}\\
\ell & \text{triv}\\
\ell-1/2& \text{def}\,.
\end{cases}
\label{AverageCharge}
\end{equation}
The SRPFs are given by 
\begin{equation}
\mathcal{Z}^{\text{triv}}_{n}(q_A)=\begin{cases}
1 & \text{for } q_A=\ell\\
0 & \text{otherwise}
\end{cases}
\end{equation}
for the trivial phase,
\begin{equation}
\mathcal{Z}^{\text{top}}_{n}(q_A)=\begin{cases}
\frac{1}{2^{2n}} & \text{for } q_A=\ell-1\\
\frac{1}{2^{2n-1}} & \text{for } q_A=\ell\\
\frac{1}{2^{2n}} & \text{for } q_A=\ell+1\\
0 & \text{otherwise}
\end{cases}
\end{equation}
for the topological phase, and 
\begin{equation}
\mathcal{Z}_{n}^{\text{def}}(q_A)=\begin{cases}
\frac{1}{2^{n}} & \text{for } q_A=\ell-1\\
\frac{1}{2^{n}} & \text{for } q_A=\ell\\
0 & \text{otherwise}\,,
\end{cases}
\end{equation}
when the defect is contained in the interval. In particular, substituting $n=1$ in the SRPFs, we obtain the probability distributions for the expectation value of the restricted charge operator $\hat{Q}_A$.  Namely, in the trivial phase $q_A=\ell$ with no other possibilities, while in the topological phase $q_A=\ell$ with probability $1/2$ and $q_A=\ell\pm1$ with probability $1/4$ each. For the case of a defect without an occupied zero mode, we find that $q_A=\ell,\ell-1$ with equal probability $1/2$. 

Combining these together and using Eq.~\eqref{eq:RE}, we find the symmetry-resolved entropies:
\begin{equation}
S^{\text{triv}}_{n}(q_A)=
0 ~~ \forall q_A
\end{equation}
for the trivial phase, 
\begin{equation}
S^{\text{top}}_{n}(q_A)=\begin{cases}
\log2 & \text{for } q_A=\ell\\
0 & \text{otherwise}
\end{cases}
\end{equation}
for the topological phase, and finally 
\begin{equation}
S_{n}^{\text{def}}(q_A)=
0 ~~ \forall q_A\,.
\end{equation}
Based on these formulas, it can be seen how the SREs for the interval give a refined information on the topology and on the defect-content of the subsystem. Indeed we have arrived at a somewhat unexpected result: whereas to calculate the unresolved entropies we only count the cut bonds and we obtain that the defect result is half of the topological result, this is not the case for SREs since $S_{n}^{\text{def}}\!\left(\ell\right)=
0$ while $S_{n}^{\text{top}}\!\left(\ell\right)=\log 2$.  

\begin{figure}[t]
\centering
\includegraphics[width=\columnwidth]{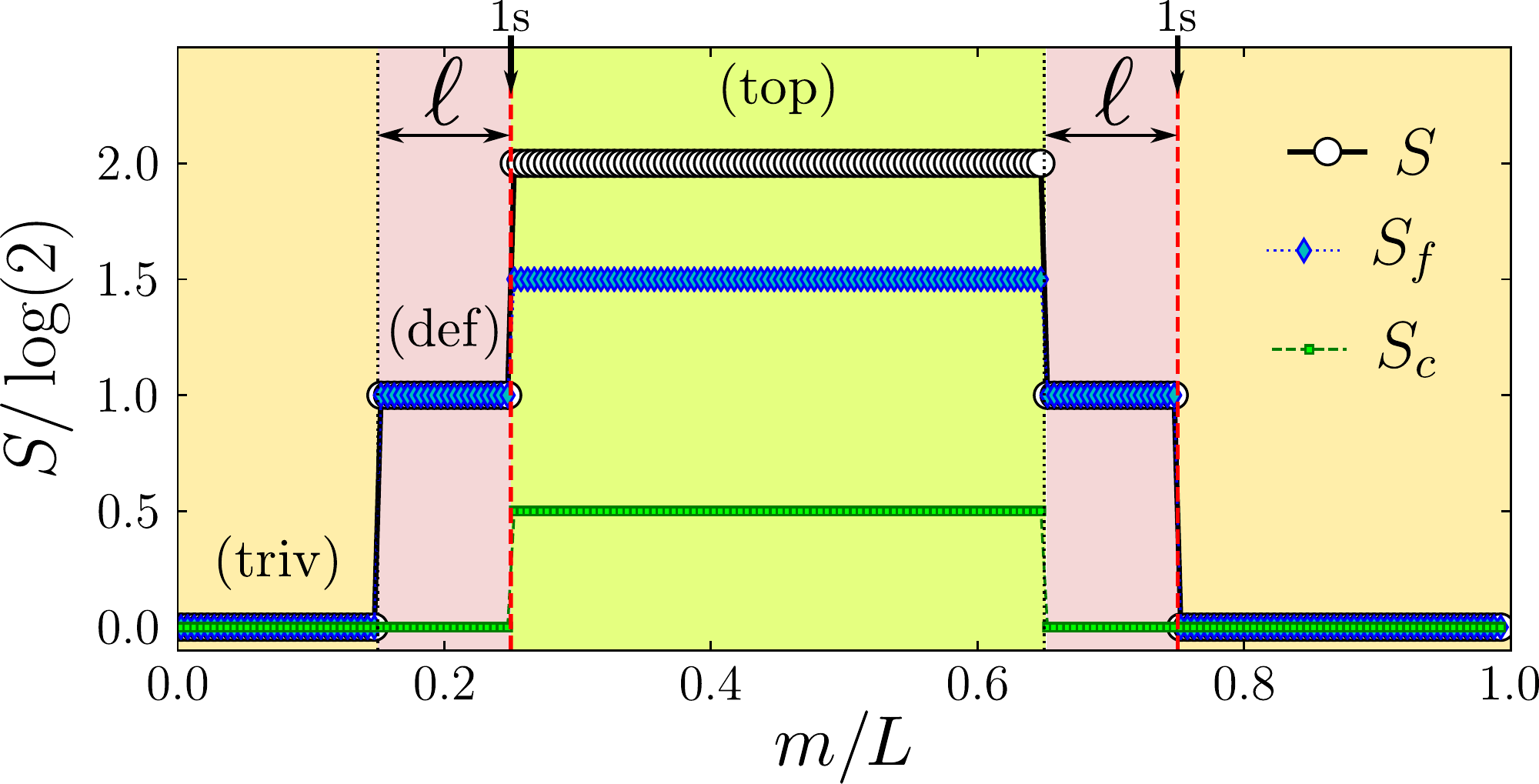}
\caption{\label{figSREPBC2kinksDimerised} Total ($S$), configuration ($S_c$) and fluctuation ($S_f$) entropies for the interval $A$ in Eq.~\eqref{eq:interval-convention}, shown as function of the interval position $m$ along the chain. We consider the fully dimerized limit $\delta=1$ of a SSH chain of length $N=2L=400$ and hopping $t=1$, and we set the subsystem size to $\ell=20$. PBC are imposed such that the model contains two 1s defects at distance $L$ (red dashed vertical lines). The system has $L-1$ excitations, such that the two-fold degenerate zero modes are not occupied.}
\end{figure}

To understand this further, we specialize to the von Neumann entropy and we recall its decomposition given in Eq.~\eqref{eq:SfSc}. From this we see how the fluctuation and configuration entropies give different contributions and play a different role in the make-up of the total entropy. Specifically,
\begin{equation}
S^{\text{triv}}_f=S^{\text{triv}}_c=0
\end{equation}
for the trivial phase, 
\begin{equation}
\begin{split}
S^{\text{top}}_{f}&=\frac{3}{2}\log2\\
S^{\text{top}}_{c}&=\frac{1}{2}\log2
\end{split}
\end{equation}
for the topological phase, and finally 
\begin{equation}
\begin{split}
S^{\text{def}}_{f}&=\log2\\
S^{\text{def}}_{c}&=0\,.
\end{split}
\end{equation}
In particular, the configuration entropy is non-vanishing only in the topological phase of the model. This means that the non-vanishing value of the total entropy with the defect must originate from the fluctuation part only. These findings are confirmed by the numerical results shown in Fig.~\ref{figSREPBC2kinksDimerised}.  

\section{Away from the fully dimerized limit of the SSH chain}
\label{AwayFromDimerized}
We now move on to considering the SSH model when the Hamiltonian of the model \eqref{SSHHamPBC} involves generic values of the dimerization parameter $\delta$, away from the fully dimerized limit. Without loss of generality, we restrict our discussion to the regime $0<\delta<1$. First, in Sec. \ref{AwayFromDimerized_AnalyticalResults}, we derive analytical expressions for the charged moments and SREs, and test them against numerical results. Aiming to put these results in a more general context, in Sec. \ref{AwayFromDimerized_StatMech} we present an argument motivated by a statistical mechanics perspective, which relates general properties of the entanglement spectrum to the observation of entanglement equipartition or its breakdown. 

\subsection{Charged moments and SREs: Analytical results}
\label{AwayFromDimerized_AnalyticalResults}

In this subsection, we derive asymptotically exact formulas for the charged moments and for the SREs of the interval $A$ defined in Eq.~\eqref{eq:interval-convention}, valid when the length $\ell$ (in terms of unit cells) of $A$ is
large. The system is again slightly below half filling, that is, zero modes associated with defects are not yet excited. Our derivation is based on exact results for the entanglement spectrum \cite{Eisler_2020} which have been obtained for half-infinite subsystems via corner transfer matrix methods. 
Similar to the unresolved entropy calculation performed in Ref.~\cite{Eisler_2020}, here the charged moment for a finite interval is readily obtained by combining the contributions from the two edges of the interval, where the contribution of each edge is related to the charged moment of a half-infinite subsystem embedded in an infinite SSH chain. The legitimacy of this procedure is based on the gapped nature of the model and on the consequence that the entanglement quickly saturates to a constant with increasing $\ell$.
Accordingly, for a large subsystem, entanglement is attributed to the two edges, whose contributions are independent from one another when $\ell\to\infty$. It also follows that, for finite subsystems, the following results are exponentially accurate due to the finite correlation length $\Delta$ of the ground state.

The essential steps in our derivation are thus to first identify the individual contributions of cut weak and strong bonds, appropriately combine them, and finally attribute the obtained quantities to intervals in the topological and in the trivial phases of the model, or to intervals containing a defect (hence involving spatial regions both in the topological phase and in the trivial phase).

Given the single-particle eigenvalues $\left\{ \epsilon_{i}\right\}$ of the entanglement Hamiltonian associated with $A$, the charged moments correspond to
\begin{equation}
Z_{n}\!\left(\alpha\right)=\prod_{i}\frac{\left(1+e^{-n\epsilon_{i}+i\alpha}\right)}{\left(1+e^{-\epsilon_{i}}\right)^{n}},
\label{ChargedMomentFromSpectrum}
\end{equation}
as can be readily verified from Eqs.~\eqref{PeschelFormula} and \eqref{Z_ChargedMomentsCf}. The question of calculating the charged moments therefore boils down to an exact computation of the single-particle entanglement spectrum together with an asymptotic analysis of the product appearing in Eq.~\eqref{ChargedMomentFromSpectrum}.

In Ref.~\cite{Eisler_2020} it was shown that the entanglement spectrum for a large but finite interval can be expressed as the direct combination of the entanglement spectra of two half-infinite subsystems, namely the subsystem obtained by fixing the right boundary of the interval and taking the left boundary to $-\infty$ and the subsystem obtained by fixing the left boundary and taking the right one to $\infty$. In other words, the charged moment can be decomposed into two independent contributions arising from its left and right boundaries, $Z_{n}\!\left(\alpha\right)=Z^{\rm{left}}_{n}\!\left(\alpha\right)Z^{\rm{right}}_{n}\!\left(\alpha\right)$, and each of the two contributions can be expressed through Eq.~\eqref{ChargedMomentFromSpectrum} by plugging in the spectrum associated with the appropriate half-infinite subsystem.

The entanglement spectrum for a half-infinite subsystem can obviously depend only on the type of bond that is cut by its boundary, meaning that it comes in two possible forms. In particular, for a cut at a strong bond we have the (non-degenerate) spectrum \cite{Eisler_2020}
\begin{equation}
\epsilon_{l}=2l\epsilon,\,\,\,\,\,l=0,\pm1,\pm2,\ldots,
\label{ModHEigvals1}
\end{equation}
while for a cut at weak bond we have
\begin{equation}
\epsilon_{l}=\left(2l-1\right)\epsilon,\,\,\,\,\,l=0,\pm1,\pm2,\ldots,
\label{ModHEigvals2}
\end{equation}
again with no degeneracy. The level separation is set by $\epsilon=\pi I\!\left(k'\right)/I\!\left(k\right)$, where $k=\left(1-\delta\right)\!/\!\left(1+\delta\right)$ and $k'=\sqrt{1-k^{2}}$, and where $I\!\left(k\right)$ is the complete elliptic integral of the first kind,
\begin{equation}
I\!\left(k\right)=\int_{0}^{1}\frac{{\rm d}x}{\sqrt{\left(1-x^{2}\right)\left(1-k^{2}x^{2}\right)}}.
\end{equation}
In what follows, we will also use the Jacobi theta functions \cite{whittaker_watson_1996} $\theta_{2}\!\left(\omega|\zeta\right)$ and $\theta_{3}\!\left(\omega|\zeta\right)$ (their explicit definitions are given in Appendix~\ref{AnalyticalChargedMomentsAppendix}), as well as the notation $\theta_{j}\!\left(\zeta\right)=\theta_{j}\!\left(0|\zeta\right)$.

In Appendix~\ref{AnalyticalChargedMomentsAppendix}, we derive the contribution of each type of boundary to the overall charged moment of the interval, using the entanglement spectra given in Eqs.~\eqref{ModHEigvals1} and \eqref{ModHEigvals2}. The contribution of a boundary located at a strong bond is found to be
\begin{equation}
Z^{\text{strong}}_{n}\!\left(\alpha\right)\propto\frac{1}{2^{\left(n-1\right)/3}}\left(\frac{{k'}^{n}}{k_{n}k_{n}'k^{2n}}\right)^{1/6}\frac{\theta_{2}\!\left(\frac{\alpha}{2}|e^{-n\epsilon}\right)}{\theta_{3}\!\left(e^{-n\epsilon}\right)},
\label{StrongBondMomentContribution}
\end{equation}
while the contribution of a boundary located at a weak bond is
\begin{equation}
Z^\text{weak}_{n}\!\left(\alpha\right)\propto\frac{1}{2^{\left(n-1\right)/3}}\left(\frac{k^{n}{k'}^{n}}{k_{n}k_{n}'}\right)^{1/6}\frac{\theta_{3}\!\left(\frac{\alpha}{2}|e^{-n\epsilon}\right)}{\theta_{3}\!\left(e^{-n\epsilon}\right)},
\label{WeakBondMomentContribution}
\end{equation}
where we defined $k_{n}$ and $k_{n}'=\sqrt{1-{k_{n}}^{2}}$ through the relation $n\epsilon=\pi I\!\left(k_{n}'\right)/I\!\left(k_{n}\right)$. In both cases, the proportionality constant is a phase factor of the form $e^{i\alpha\Gamma}$, where $\Gamma$ effectively counts the number of negative pseudo-energies of the entanglement Hamiltonian, and is thus related to the finite length $\ell$ of the subsystem (see Appendix~\ref{AnalyticalChargedMomentsAppendix}). This phase factor is therefore ill-defined when considering the independent contribution of each boundary, but its overall contribution to the charged moment can be determined simply by observing that it sets the value of the average charge in the subsystem.

More precisely, in order to determine the charged moment of the subsystem, we need to multiply the two boundary contributions (given by Eqs.~\eqref{StrongBondMomentContribution} and \eqref{WeakBondMomentContribution}) according to the type of bonds that these boundaries cut, and then multiply this product by $\exp\left(i\alpha\langle \hat{Q}_A\rangle\right)$ to obtain the exact asymptotic result. The average charge $\langle \hat{Q}_A\rangle$ is given in Eq.~\eqref{AverageCharge}. This procedure can also be seen as one that arises from the requirement that the following results for the charged moments converge to those derived for the fully dimerized limit (Eq.~\eqref{FullyDimerizedChargedMoments}) in the limit $\delta\to1$ (corresponding to $\epsilon\to\infty$).

In total, when the subsystem does not contain a defect, for the trivial and the topological phase the calculation simply amounts to squaring the boundary contribution of a weak bond or a strong bond (respectively), and setting the multiplicative phase factor to $e^{i\alpha\ell}$. If the subsystem does contain a defect, we need to multiply the contributions of a strong bond and a weak bond, with the phase factor set to $e^{i\alpha(\ell-1/2)}$. Therefore, we obtain
\begin{equation}
\begin{split}
Z^\text{top}_{n}\!\left(\alpha\right)=&\,\frac{e^{i\alpha\ell}}{4^{\left(n-1\right)/3}}\left(\frac{{k'}^{n}}{k_{n}k_{n}'k^{2n}}\right)^{1/3}\left[\frac{\theta_{2}\!\left(\frac{\alpha}{2}|e^{-n\epsilon}\right)}{\theta_{3}\!\left(e^{-n\epsilon}\right)}\right]^{2},\\
Z^\text{triv}_{n}\!\left(\alpha\right)=&\,\frac{e^{i\alpha\ell}}{4^{\left(n-1\right)/3}}\left(\frac{k^{n}{k'}^{n}}{k_{n}k_{n}'}\right)^{1/3}\left[\frac{\theta_{3}\!\left(\frac{\alpha}{2}|e^{-n\epsilon}\right)}{\theta_{3}\!\left(e^{-n\epsilon}\right)}\right]^{2},\\
Z^\text{def}_{n}\!\left(\alpha\right)=&\,\frac{e^{i\alpha(\ell-1/2)}}{4^{\left(n-1\right)/3}}\!\!\left(\!\frac{{k'}^{n}}{k_{n}k_{n}'k^{n/2}}\!\right)^{\!\!\!1/3}\\
&\times\frac{\theta_{2}\!\left(\frac{\alpha}{2}|e^{-n\epsilon}\right)\theta_{3}\!\left(\frac{\alpha}{2}|e^{-n\epsilon}\right)}{\left[\theta_{3}\!\left(e^{-n\epsilon}\right)\right]^{2}}.
\end{split}
\label{ExactChargedMoments}
\end{equation}
A noteworthy observation is that the logic leading to the formula for charged moments for an interval containing a defect also implies that the corresponding entanglement spectrum, in the asymptotic limit, can be simply written as
\begin{equation}
\epsilon_{l}=l\epsilon,\,\,\,\,\,l=0,\pm1,\pm2,\ldots,
\label{ModHEigvalsDef}
\end{equation}
as it is a direct combination of the spectra in Eqs.~\eqref{ModHEigvals1} and \eqref{ModHEigvals2}. The spectrum in Eq.~\eqref{ModHEigvalsDef} is a simple modification of Eq.~\eqref{ModHEigvals1}, the spectrum arising from a \emph{single} boundary at a strong bond. This finding is also confirmed numerically in Sec. \ref{SectionWithZM}.  

We can now determine the SREs away from the dimerized limit by performing the Fourier transform of the charged moments in Eq.~\eqref{ExactChargedMoments}. In Appendix~\ref{AnalyticalSRPFAppendix}, we explain how the Fourier transform of Eq.~\eqref{eq:CalU1} can be computed analytically in this case. Defining $\Delta q = q_A - \ell$, for both the topological and the trivial cases, we find that the SRPFs depend only on the parity of $\Delta q$, apart from an overall Gaussian multiplicative factor (centered at $\Delta q =0$). Namely, our computation yields
\begin{equation}
\begin{split}
\mathcal{Z}_n^{\text{top}}\!\left(q_A\right)=&\,\frac{e^{-\frac{1}{2}n\epsilon\,\left(\Delta q\right)^2}}{4^{\left(n-1\right)/3}\left[\theta_{3}\!\left(e^{-n\epsilon}\right)\right]^2}\left(\frac{{k'}^{n}}{k_{n}k_{n}'k^{2n}}\right)^{1/3} \\
&\times\begin{cases}
\theta_3\!\left(e^{-2n\epsilon}\right) & \Delta q {\text{ is odd}} \\
\theta_2\!\left(e^{-2n\epsilon}\right) & \Delta q {\text{ is even}}\,,
\end{cases}\\
\mathcal{Z}_n^{\text{triv}}\!\left(q_A\right)=&\,\frac{e^{-\frac{1}{2}n\epsilon\,\left(\Delta q\right)^2}}{4^{\left(n-1\right)/3}\left[\theta_{3}\!\left(e^{-n\epsilon}\right)\right]^2}\left(\frac{k^{n}{k'}^{n}}{k_{n}k_{n}'}\right)^{1/3} \\
&\times\begin{cases}
\theta_2\!\left(e^{-2n\epsilon}\right) & \Delta q {\text{ is odd}} \\
\theta_3\!\left(e^{-2n\epsilon}\right) & \Delta q {\text{ is even}}\,.
\end{cases}
\end{split}
\label{AnalyticalSRPF_TopTriv}
\end{equation}
In contrast, when the subsystem contains a defect, the sectors do not split according to their parity, and the SRPFs are given by 
\begin{equation}
\mathcal{Z}_n^{\text{def}}\!\left(q_A\right)=\,\frac{e^{-\frac{1}{2}n\epsilon\,\left(\Delta q+\frac{1}{2}\right)^2}\theta_2\!\left(e^{-n\epsilon/2}\right)}{4^{\left(2n+1\right)/6}\left[\theta_{3}\!\left(e^{-n\epsilon}\right)\right]^2}\left(\frac{{k'}^{n}}{k_{n}k_{n}'k^{n/2}}\right)^{1/3}.
\label{AnalyticalSRPF_Def}
\end{equation}
The SRPF therefore simply corresponds to a Gaussian distribution function centered at $\Delta q = -\frac{1}{2}$.

The results of Eqs.~\eqref{AnalyticalSRPF_TopTriv} and \eqref{AnalyticalSRPF_Def} allow us to directly write down the R\'enyi SREs for all cases:
\begin{equation}
\begin{split}
S_n^{\text{top}}\!\left(q_A\right)&=\sigma_n+\frac{1}{1-n}\begin{cases}
\log\!\left[\frac{\theta_3\!\left(e^{-2n\epsilon}\right)}{\left[\theta_3\!\left(e^{-2\epsilon}\right)\right]^n}\right] & \Delta q {\text{ is odd}} \\
\log\!\left[\frac{\theta_2\!\left(e^{-2n\epsilon}\right)}{\left[\theta_2\!\left(e^{-2\epsilon}\right)\right]^n}\right] & \Delta q {\text{ is even}}\,,
\end{cases}\\
S_n^{\text{triv}}\!\left(q_A\right)&=\sigma_n+\frac{1}{1-n}\begin{cases}
\log\!\left[\frac{\theta_2\!\left(e^{-2n\epsilon}\right)}{\left[\theta_2\!\left(e^{-2\epsilon}\right)\right]^n}\right] & \Delta q {\text{ is odd}} \\
\log\!\left[\frac{\theta_3\!\left(e^{-2n\epsilon}\right)}{\left[\theta_3\!\left(e^{-2\epsilon}\right)\right]^n}\right] & \Delta q {\text{ is even}}\,,
\end{cases}\\
S_n^{\text{def}}\!\left(q_A\right)&=\sigma_n+\frac{1}{1-n}\log\!\left[\frac{2^{n-1}\theta_2\!\left(e^{-n\epsilon/2}\right)}{\left[\theta_2\!\left(e^{-\epsilon/2}\right)\right]^n}\right],
\label{ExactSREs}
\end{split}
\end{equation}
where we introduced the notation
\begin{equation}
    \sigma_n=\frac{1}{1-n}\log\!\left[\frac{\left[\theta_{3}\!\left(e^{-\epsilon}\right)\right]^{2n}}{\left[\theta_{3}\!\left(e^{-n\epsilon}\right)\right]^2}\left(\frac{k^n{k'}^{n}}{4^{n-1}k_{n}k_{n}'}\right)^{1/3}\right].
\end{equation}
An analytic continuation to $n=1$, which produces the von Neumann SRE, can the in principle be done using the identity $k_n=\left[\theta_{2}\!\left(e^{-n\epsilon}\right)/\theta_{3}\!\left(e^{-n\epsilon}\right)\right]^2$ \cite{whittaker_watson_1996}.

Based on the exact results for the SREs in Eq. \eqref{ExactSREs}, we observe the emergence of novel features that were absent in the $\delta=1$ case, since more than one charge sector has non-vanishing contributions.  In particular, an interesting connection can be made between the topological phase and the trivial phase. In the trivial phase, we find that charge sectors with $\langle \hat{Q}_A\rangle\pm (2q-1)$ for $q\in \mathbb{N}$ all have an equal contribution, which is much larger than the equal contribution of sectors with $\langle \hat{Q}_A\rangle\pm 2q$, while in the topological case the opposite is true.
That is, we can observe that, in the topological and trivial phases, the overall entanglement equipartition is broken, but there is exact equipartition within each charge-parity sector. Moreover, the value of an even-charge SRE in the topological phase is equal to that of an odd-charge SRE in the trivial phase, and vice versa. Interestingly, in the presence of the defect, all charge sectors have an equal contribution, that is, the conventional equipartition of entanglement is found. This fact is worth stressing, since it means that for the SSH model, the entanglement equipartition in the usual sense is only present if the subsystem includes a defect (as long as the associated zero mode is not excited).

\begin{figure}[t]
    \centering
    \includegraphics[width=\columnwidth]{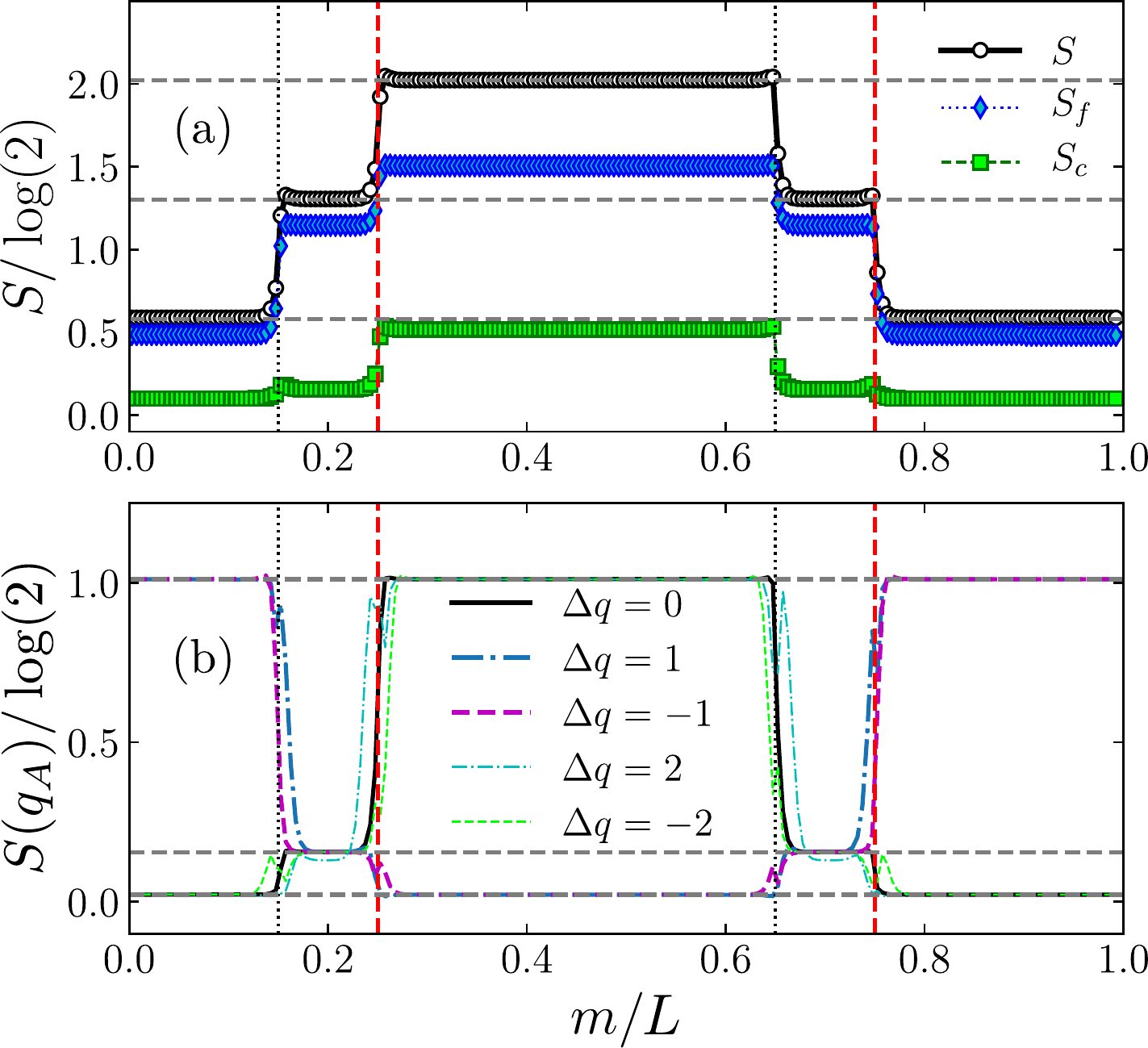}
\caption{\label{figSREPBC2kink1s1s} {\it Panel} (a)~--~ Total ($S$), configuration ($S_c$) and fluctuation ($S_f$) entropies for the interval $A$ in Eq.~\eqref{eq:interval-convention}, shown as function of the interval position $m$ along the chain. We set up the system as in Figure~\ref{figSREPBC2kinksDimerised} but with dimerization parameter $\delta=0.3$. {\it Panel} (b)~--~Symmetry-resolved entanglement for different values of charge $\Delta q=q_A-\ell$ for the interval $A$ in Eq.~\eqref{eq:interval-convention}, shown as function of the interval position $m$ along the chain. In both panels, horizontal dashed lines show the asymptotically exact values of the entropies. The red dashed vertical lines mark the location of the 1s defects.}
\end{figure}

The exact results for the SREs in Eq.~\eqref{ExactSREs} are visualized in Fig.~\ref{figSREPBC2kink1s1s}. We cross-checked these results, along with the results for the total entropies (obtained from~\eqref{ExactChargedMoments} at $\alpha=0$ and from \eqref{TotalRenyi} and \eqref{TotalVN}), against exact numerical computations of the correlation matrix after determining first the single-particle eigenstates of the Hamiltonian numerically and applying Eq.~\eqref{CijSumk}.  As shown in  Fig.~\ref{figSREPBC2kink1s1s}, the agreement is excellent thereby justifying our expressions for the charge moments and the SREs. For a simplified presentation, we only display the analytical predictions for the entropies,  although our analytical results nicely agree with the numerically obtained probability distribution as well.

From these results, we once again observe a finer structure of the entanglement which is revealed when considering symmetry resolution.  In particular,  the total entropy of the interval with a defect is again simply the sum of the contributions of a weak and a strong bond.  However,  we observe that, as in the fully dimerized case, the entropy redistributes between its configuration and fluctuation parts in the different regimes. Indeed, even though for $0<\delta<1$ we find a non-vanishing configuration entropy when the defect is included in the interval, its value is strongly suppressed compared to the fluctuation entropy. This suppression is more significant in the trivial phase and it turns out to give a considerable contribution only in the topological phase.

Additionally, we mention that the behavior of the SRPFs $\mathcal{Z}_n\!\left(q_A\right)$ remains qualitatively very similar to the case of the fully dimerized limit. In particular, the probabilities of the $\ell \pm \Delta q$ charge components without the defect are strongly suppressed for $\Delta q\geq2$, although non-zero. Similarly, in the presence of the defect, the dominant contributions are attributed to the sectors with charge $\ell$ and $\ell-1$.

\subsection{Statistical mechanics perspective on entanglement equipartition}
\label{AwayFromDimerized_StatMech}

It is illuminating to note that the source of the parity patterns that were observed in Sec.~\ref{AwayFromDimerized_AnalyticalResults} can be understood through a relatively simple statistical mechanics argument, applied to the entanglement Hamiltonian and to its single-particle pseudo-energies. The basic rationale behind this argument is that the projection of the entanglement Hamiltonian onto a specific charge sector may be regarded as the imposition of a constraint on the \emph{average} charge, provided that the ensuing charge fluctuations are very small.

This constraint amounts to replacing the entanglement Hamiltonian $\hat{H}_A=-\log \rho_A$ with $\hat{H}_A-\mu\hat{Q}_A$, where $\mu$ is the Lagrange multiplier associated with the constraint. This yields the modified RDM
\begin{equation}
\widetilde{\rho}_A=\frac{1}{\mathcal{N}_{\mu}}\exp\!\left(-\hat{H}_A+\mu\hat{Q}_A\right),
\end{equation}
where $\mathcal{N}_{\mu}={\rm Tr}\!\left[e^{-\hat{H}_A+\mu\hat{Q}_A}\right]$. $\mu$ can be thought of as a fictitious chemical potential, which induces an effective shift of the single-particle spectrum of $\hat{H}_A$. Namely, if $\left\{\epsilon_i\right\}$ is the original single-particle entanglement spectrum, then under the constraint of an average charge equal to $q_A$ it becomes $\left\{\epsilon_i-\mu\right\}$, with $\mu$ determined self-consistently through
\begin{equation}
q_A=\sum_{i=1}^{2\ell}\frac{1}{e^{\epsilon_{i}-\mu}+1}\,.
\label{ChargeConstraint}
\end{equation}

If we now let $\widetilde{S}\!\left(q_A\right)$ denote the entropy of $\widetilde{\rho}_A$, it is simply given by
\begin{equation}
\begin{split}
\widetilde{S}\!\left(q_A\right)= & -\sum_{i=1}^{2\ell}\frac{1}{e^{\epsilon_{i}-\mu}+1}\log\!\left(\frac{1}{e^{\epsilon_{i}-\mu}+1}\right) \\
& -\sum_{i=1}^{2\ell}\frac{e^{\epsilon_{i}-\mu}}{e^{\epsilon_{i}-\mu}+1}\log\!\left(\frac{e^{\epsilon_{i}-\mu}}{e^{\epsilon_{i}-\mu}+1}\right),
\end{split}
\label{EntropyWithChemicalPotential}
\end{equation}
which implies, in particular, that a certain pseudo-energy $\epsilon_{i}$ will have a larger entropy contribution if it is closer to $\mu$. As the constraint on the \emph{average} charge still allows charge fluctuations, this entropy can be decomposed into contributions from \emph{exact} charge sectors in analogy to Eq.~\eqref{eq:SfSc}. That is, we may write
\begin{equation}
\widetilde{S}\!\left(q_A\right)=\sum_{q_{A}'}\mathcal{Z}_1'\!\left(q_{A}'\right)S\!\left(q_{A}'\right)-\sum_{q_{A}'}\mathcal{Z}_{1}'\!\left(q_{A}'\right)\log \mathcal{Z}_{1}'\!\left(q_{A}'\right)\,.
\label{ConstrainedEntropyDecomposition}
\end{equation}
Here the primed notation $\mathcal{Z}_1'$ emphasizes that the charge distribution is not the same as in Eq.~\eqref{eq:SfSc}, but is rather modified according to the average charge constraint.

Crucially, the SREs $S\!\left(q_{A}'\right)$ in Eq.~\eqref{ConstrainedEntropyDecomposition} are the same as those that appear in Eq.~\eqref{eq:SfSc}. This can be seen by considering two (many-body) eigenstates of $\hat{H}_A$, $|\phi_1\rangle$ and $|\phi_2\rangle$, that are within the same charge sector, with charge $q_{A}'$. The probabilities associated with these eigenstates then satisfy
\begin{equation}
\langle\phi_j|\widetilde{\rho}_A|\phi_j\rangle=\frac{e^{\mu q_A'}}{\mathcal{N}_{\mu}}\langle\phi_j|\rho_A|\phi_j\rangle\,,
\end{equation}
meaning that the chemical potential affects the probabilities of states within the same charge sector by multiplying them by the same factor, thus making the distribution (and hence also the entropy) within each charge sector independent of $\mu$.

Note that, in Eq.~\eqref{ConstrainedEntropyDecomposition}, the probability $\mathcal{Z}_1'\!\left(q_{A}\right)$ for measuring the average charge should typically be dominant among the modified charge distribution, thus facilitating the extraction of $S\!\left(q_{A}\right)$ from Eq.~\eqref{ConstrainedEntropyDecomposition}, provided that one has proper knowledge of all other terms appearing there. This becomes particularly simple if charge fluctuations are negligible, entailing that $\mathcal{Z}_1'\!\left(q_{A}\right)\approx1$. This requires that $\mu$ would be sufficiently far from any pseudo-energy level $\epsilon_i$, so that any such level can be approximately regarded as either fully occupied (if $\mu>\epsilon_i$) or completely empty (if $\mu<\epsilon_i$).

To showcase the strength of this statistical mechanics picture, let us consider first the case of the trivial phase, where the entanglement spectrum is the combination of two identical edge contributions, given by Eq.~\eqref{ModHEigvals2}. This doubly-degenerate spectrum is centered around 0, and it is easy to check that in the case $q_A=\ell$, Eq.~\eqref{ChargeConstraint} necessarily gives $\mu=0$. Thus, the chemical potential $\mu$ resides at the middle of a gap of size $2\epsilon$ in the spectrum. For larger $\epsilon$ (which is equivalent to $\delta$ being closer to $1$), the entropy contribution of each pseudo-energy level diminishes according to Eq.~\eqref{EntropyWithChemicalPotential}, yet charge fluctuations also become more suppressed and thus the approximation $S\!\left(q_A=\ell\right)\approx\widetilde{S}\!\left(q_A=\ell\right)$ for the SRE becomes exact as $\delta\to1$.

In contrast, if we choose to set the average charge to $q_A=\ell+1$, then Eq.~\eqref{ChargeConstraint} states that $\mu$ should coincide with the doubly-degenerate level $\epsilon_1$, that is, $\mu=\epsilon$. The two levels induce significant charge fluctuations, as each of them is occupied with probability $1/2$. Assuming that all levels other than the two at $\epsilon_1$ are either almost empty or almost full (which becomes exact as $\delta\to1$), we have $\mathcal{Z}_1'\!\left(\ell+1\right)\approx1/2$ and $\mathcal{Z}_1'\!\left(\ell\right)\approx\mathcal{Z}_1'\!\left(\ell+2\right)\approx1/4$, while the latter two charge labels are associated with negligible entropy. Plugging this into Eq.~\eqref{ConstrainedEntropyDecomposition}, we find
\begin{equation}
S\!\left(q_A=\ell+1\right)\approx2\widetilde{S}\!\left(q_A=\ell+1\right)-3\log2\,.
\label{ConstrainedEntropyFluctuations}
\end{equation}
In particular, since $\widetilde{S}\!\left(q_A=\ell+1\right)\approx2\log2$ given that the two levels coinciding with $\mu$ have a maximal entropy contribution, we observe that $S\!\left(q_A=\ell+1\right)\approx\log2$, an enhanced value of the entropy compared to $S\!\left(q_A=\ell\right)$. Note that in Fig.~\ref{figSREPBC2kink1s1s}, the value of $S\!\left(q_A=\ell\pm1\right)$ in the regions corresponding to the trivial phase is indeed very close to $\log2$, even though it was computed for $\delta=0.3$, i.e., not particularly close to the fully dimerized limit.

Since the entire single-particle spectrum is doubly-degenerate, it is now straightforward to deduce the general rule stating that $S\!\left(q_A\right)$ is enhanced for odd values of $\Delta q$ and diminished for even values of $\Delta q$. Furthermore, given that the spectrum is also equidistant (referring to the distance between degenerate pairs), indeed the entropies should be equal within each parity sector, in agreement with our analytical result in Eq.~\eqref{ExactSREs}. The last observation is not exact for finite $\ell$ even if we assume that the finite entanglement spectrum remains equidistant (e.g., for even $\Delta q\neq0$, the average charge constraint will not fix $\mu$ exactly at the center of the gap as it does for $\Delta q=0$), but these effects arising from the spectrum edges should be very weak, by the same reasoning that allowed us to compute the finite-interval spectrum from the spectra of half-infinite subsystems.

This argument applies by extension to the case of the topological phase, when considering that the doubly-degenerate entanglement spectrum of Eq.~\eqref{ModHEigvals1} is centered around two levels at $\epsilon_0=0$, which both have a maximal entropy contribution for $q_A=\ell$, as the corresponding chemical potential is set to $\mu=0$. The same logic that was applied for the trivial phase then dictates the opposite dependence on charge parity for the topological phase. For the case where the subsystem contains a defect, the exact equipartition simply stems by the same logic from the fact that the entanglement spectrum $\left\{\epsilon_l\right\}$ of Eq.~\eqref{ModHEigvalsDef} is non-degenerate and equidistant: for each choice of $q_A$, $\mu$ is set inside a different gap in the spectrum, but the shifted spectrum $\left\{\epsilon_l-\mu\right\}$ is always the same.

\section{Exciting the zero modes of the defects}
\label{SectionWithZM}

In the previous sections, we focused on cases where the zero modes localized at the defects were not included in the ground state, that is, the system was slightly below half filling. Here, we elaborate on the behavior of entropies when zero modes are excited as well. The quantities we computed so far remain essentially unaltered if the interval does not include a defect supporting an excited zero mode. This is a consequence of the fact that the zero modes are always exponentially localized and hence if the interval making up the subsystem is far away from the defects, the impact of a zero mode is exponentially small.

The main focus of our discussion below will therefore be on the case where the subsystem contains a defect. The situation remains relatively simple when the zero mode associated with the defect is fully occupied.
In the fully dimerized limit, all the formulas presented in Sec.~\ref{SREFullyDimerised} remain unchanged, apart from those regarding the case of the defect: here one simply has to substitute $\ell\rightarrow \ell+1$ in the equations of Sec.~\ref{SREFullyDimerised}. Since the formulas for the charged moments and partition functions (including the charge probability distribution) undergo a simple shift in the charge sectors labeling, it follows that the total and symmetry-resolved entropies remain unaltered. 

This statement holds true even away from the fully dimerized limit, apart from exponentially small spatial regions in which the support of the zero modes are split, so that they have weight both inside and outside the interval. Apart from this, once again the substitution $\ell\rightarrow \ell+1$ in $Z_n^\text{def}(\alpha)$ appearing in Eq.~\eqref{ExactChargedMoments} gives the correct result. The total and resolved entropies remain again unaltered.\\

The situation becomes more intricate when the chain hosts two defects -- one inside the subsystem and the other outside of it -- and only a single zero mode is excited. A crucial physical feature is the fact that zero modes can be localized on several defects. 
In particular, when we consider systems with two defects
that are away from the fully dimerized limit, the two zero modes associated with the defects can be regarded as localized wave functions denoted as $|\Psi_{1}\rangle$ and $|\Psi_{2}\rangle$. They can be constructed following a similar logic to the treatment of edge modes in Ref.~\cite{Asb_th_2016}. The localization length of these wave functions is given by Eq.~\eqref{LocLength}, that is, they have a $e^{-d/\xi}$ support on the other defect at distance $d$, which is negligible for sufficiently large separation $d$, and which vanishes in the thermodynamic limit.

However, for a finite-size system, the zero modes exhibit a non-vanishing overlap with each other, and are related by a non-zero matrix element of the Hamiltonian. This means that their hybridization always yields an energetically more favorable state, and the true eigenstates are always hybridized states of $|\Psi_{1}\rangle $ and $|\Psi_{2}\rangle$. The energies of either the truly-localized or the hybridized states, however, are exactly zero in the fully dimerized limit, or away from it in the thermodynamic limit and for an infinite separation. Otherwise, in general, these energies are found to be exponentially small when $d\gg 1$. 
This means that we can eventually consider any superposition of the two zero modes, either using the truly-localized modes or the hybridized ones. Such superpositions are exact eigenstates in the thermodynamic limit and in the fully dimerized case, and they give an exponentially accurate ${\cal O}(e^{-d/\xi})$ approximation of them for large (but finite) systems with largely separated defects.

Therefore, below we investigate the impact on SREs originating from the excitation of a generic superposition of the two zero modes. In particular, we introduce the hybridization parameter $0\leq p\leq 1$ such that the (approximate) zero mode we excite is in the form
\begin{equation}
|\Psi_{p}\rangle=\sqrt{1-p}~|\Psi_{1}\rangle+e^{i\phi}\sqrt{p}~|\Psi_{2}\rangle\,.
\label{EpsPsi}
\end{equation}
The wave function $\ket{\Psi_p}$ is then a superposition of the truly-localized zero modes with probability $1-p$ on the first defect and $p$ on the second and $e^{i\phi}$ is a phase factor that has no effect on the quantities in which we are interested.
By tuning the parameter $p$, one can interpolate between a truly-localized mode ($p=0,1$) and a two-defect localized (with equal weights) state ($p=0.5$).
It is useful to clarify that in reality such hybridized zero modes, in the case of the SSH chain, are known to be sensitive to the environment and in practice only a truly localized state can be stable. Additionally, even if we assume extreme isolation of the system as well as a very fine-tuned chemical potential, preparing states like \eqref{EpsPsi} can be non-trivial: for two localized zero modes, their finite (though exponentially small) overlap will result in the true ground state having an equal support on the two defects. However, if the chain contains more than two interfaces (defects and/or edges) whose relative distances are not equal, their uneven separations can generally result in the support of the ground state on a certain defect being equal to any value between 0 and 1. In other words, the localization probability $p$ can be regarded as a truly physical parameter, at least from the viewpoint of preparing non-trivial zero mode states. For the sake of simplicity and transparency, nevertheless, we continue our investigations with a chain hosting only two defects, and regard \eqref{EpsPsi} as a given quantum state.

In the fully dimerized limit of the SSH chain, the excitation of $|\Psi_p\rangle$ leads to one zero eigenvalue of the correlation matrix $C_\text{def}$, associated with, e.g.,~defect 1, changing to $1-p$ and, accordingly, the SRPFs change to
\begin{equation}
\mathcal{Z}_{n}^{\left(p\right)}(q_A)=\begin{cases}
\frac{p^n}{2^n} & \text{for } q_A=\ell-1\\
\frac{(1-p)^n+p^n}{2^n} & \text{for } q_A=\ell\\
\frac{(1-p)^n}{2^n} & \text{for } q_A=\ell+1 \\
0 & \text{otherwise}\,,
\end{cases}
\label{2DefectFullyDimerizedSRPF}
\end{equation}
from which the charge probability distributions are easy to recover as well. The SREs read as
\begin{equation}
S_{n}^{\left(p\right)}(q_A)=\begin{cases}
\frac{1}{1-n}\log\left((1-p)^n+p^n\right) & \text{for } q_A=\ell\\
0 & \text{otherwise}\,,
\end{cases}
\end{equation}
and, taking $n\rightarrow 1$,
\begin{equation}
S^{\left(p\right)}(q_A)=\begin{cases}
-p \log p-(1-p)\log(1-p) & \text{for } q_A=\ell\\
0 & \text{otherwise}\,.
\end{cases}
\end{equation}
As we can see, the charge fluctuations behave in a natural way, and the resolved and total entropies acquire an excess contribution $\Delta S^{\left(p\right)}_n=S^{\left(p\right)}_n-S^{\left(p=1\right)}_n$ compared to the $p=1$ case (i.e., when the zero mode is empty), which reads as
\begin{equation}
\begin{split}
\Delta S^{\left(p\right)}_n&=\Delta S^{\left(p\right)}_n\left(q_A=\ell\right)=\frac{1}{1-n} \log\left((1-p)^n+p^n\right),\\
\Delta S^{\left(p\right)}&=\Delta S^{\left(p\right)}\left(q_A=\ell\right)=-p\log p-(1-p)\log(1-p).
\label{ExcessEntropySSH}
\end{split}
\end{equation}

\begin{figure}[t]
\centering
\includegraphics[width=\columnwidth]{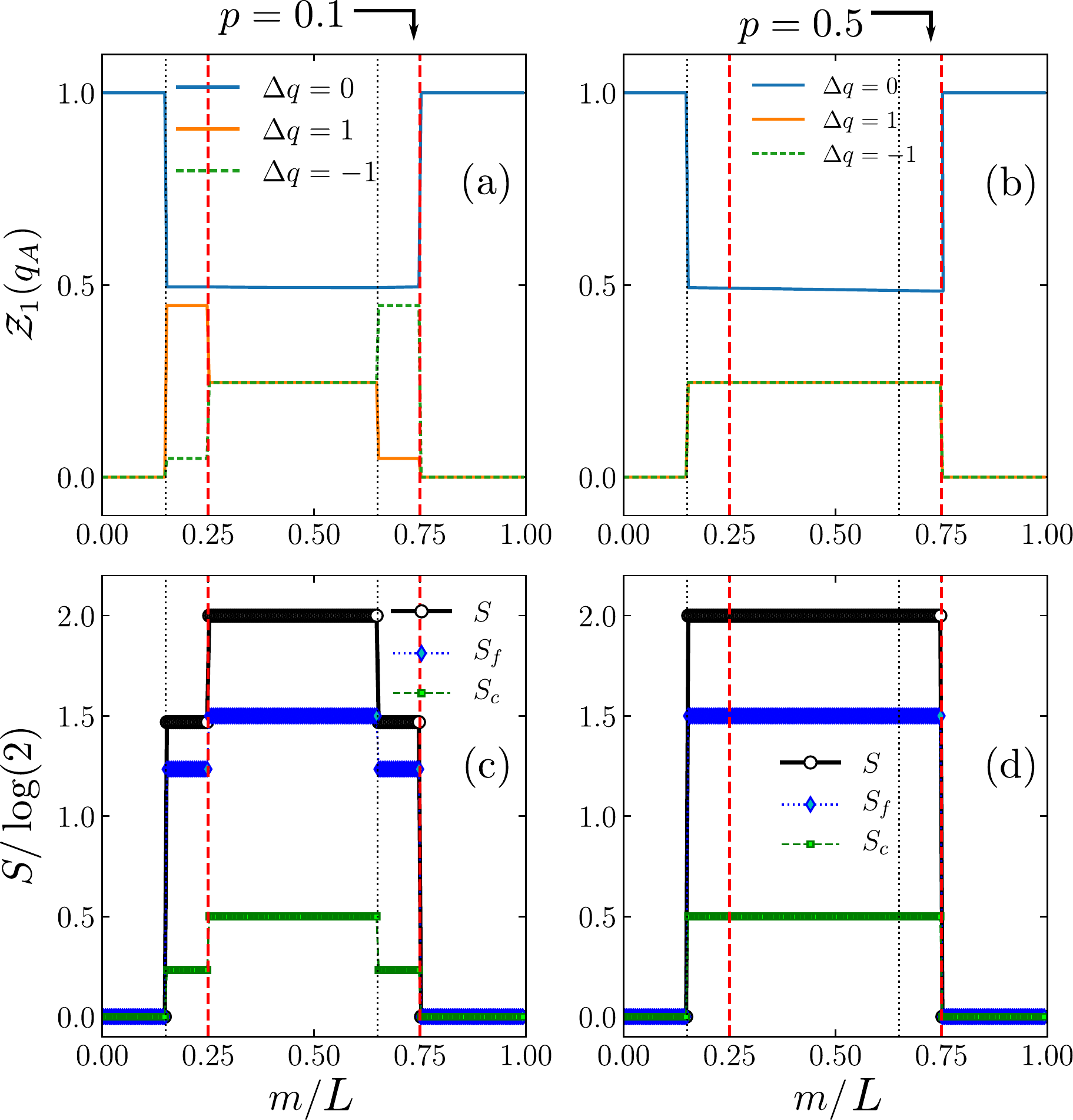}
\caption{\label{figSREPBC2kinksDimerised2} {\it Panels} (a,b)~--~The charge resolved partition functions $\mathcal{Z}_{1}(q_A)$ for the interval $A$ in Eq.~\eqref{eq:interval-convention} for different values of $\Delta q=q_A-\ell$, with $p=0.1$ (panel (a)) and $p=0.5$ (panel (b)), shown as function of the interval position $m$. {\it Panels} (c,d)~--~The corresponding total ($S$), configuration ($S_c$) and fluctuation ($S_f$) entropies. In all panels, we consider the same setup of Figure~\ref{figSREPBC2kinksDimerised}, but we excite $L$ modes such that one of the two-fold degenerate zero modes is occupied. The parameter $p$ controls the weight of the zero mode on the two 1s defects (see Eq.~\eqref{EpsPsi}).}
\end{figure}

In Fig.~\ref{figSREPBC2kinksDimerised2}, we show the numerical results for the SRPFs (panels (a,b)) and for the entropies (panels (c,d)) for two different values of $p$. Regarding the $\mathcal{Z}_1$ SRPFs, we can observe natural changes in the probabilities of the $\Delta q=\pm1$ sectors if the defect is included in the subsystem. In particular, when $p=0.5$ we can notice that the behavior of the topological phase is extended also to the region where the interval contains the defect. The features of the topological phase  originate from two cut bonds, whereas those of the interval including a defect stem from one cut bond and a "cut" zero mode, or, in other words, from the fluctuation of the two-defect localized zero mode through its non-trivial RDM as a consequence of the spatial partitioning. When the localization of the zero mode becomes asymmetric, this fluctuation contribution alters and eventually vanishes, and features similar to the trivial phases become prevalent. The behavior of the SREs is also easy to interpret. The either symmetrically or asymmetrically localized two-defect mode has an additive and independent excess entropy contribution dictated by Eq.~\eqref{ExcessEntropySSH}. This contribution affects only an interval containing the defect, that is, when the zero mode is `cut' by the spatial partitioning, and it depends on the parameter $p$ in a simple fashion (in particular, it is maximal when $p=0.5$).

Notice that, according to Eq.~\eqref{ExcessEntropySSH}, when a zero mode is close to being fully localized on one defect ($p=1$ or $p=0$), any additional contribution in the entropies vanishes. Importantly, charge sectors other than $q_A=\ell$ have no SRE contribution at all. This latter statement is in sharp contrast with the results found when the system is away from the fully dimerized limit, which we now  discuss.

We shall now demonstrate that, away from the fully dimerized limit, the SREs very sensitively detect charge imbalance on the two defects. In particular, if the zero mode wave function only mildly deviates from a truly-localized state, the SRE always has a component (up to the maximum charge value allowed in the subsystem) which is significantly enhanced, even while the total entropy changes only negligibly. This corresponds to a significant breaking of entanglement equipartition.
\begin{figure*}[t]
\includegraphics[width=\textwidth]{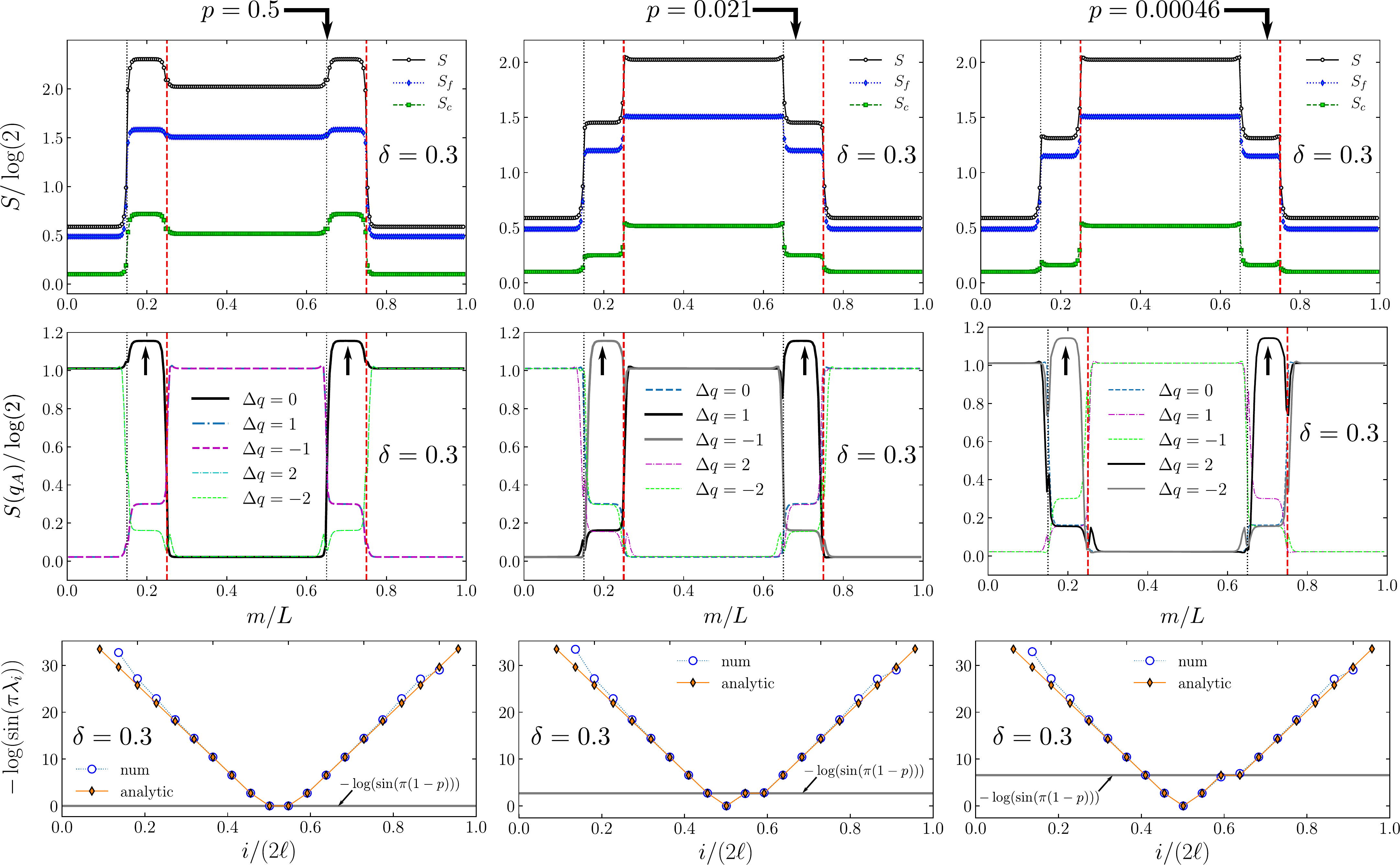}
\caption{\label{figSREPBCkink1s3sSublattice} {\it Top row}~--~Total ($S$), configuration ($S_c$) and fluctuation ($S_f$) entropies; {\it Middle row}~--~corresponding symmetry-resolved entanglement for various $\Delta q=q_A-\ell$. In both the first and the second rows, we consider a subsystem $A=[m,m+\ell-1]$ of size $\ell=20$ and position $m$ varied along the chain. The data is obtained for the SSH model with PBC, system size $N=2L=400$, hopping $t=1$, dimerization parameter $\delta=0.3$ and two 1s defects located at sites $L/4$ and $3L/4$. The spectrum contains $L$ modes, i.e., it includes a linear combination of the two zero modes localized at the defects, weighted by $p$ (c.f.~Eq.~\eqref{EpsPsi}). Different columns show the behavior upon varying of $p$. {\it Bottom row}~--~Analytical (diamonds, c.f.~ Eq.~\eqref{PeschelFormula} and \eqref{ModHEigvalsDef}) and numerical (circles) spectrum of the correlation matrix $\lambda_i$ when the interval is centered around the first defect, displayed using the function $-\log(\sin(\pi\lambda_i))$. The horizontal gray line corresponds to the analytical prediction for the eigenvalue associated with the zero mode, $-\log(\sin\pi(1-p))$.
}
\end{figure*}

More precisely, in Fig.~\ref{figSREPBCkink1s3sSublattice} we observe a systematic redistribution among charge sectors of the contribution coming from the excited zero mode $|\Psi_p\rangle$. The farther the parameter $p$ is from $1/2$, the larger is the difference $|\Delta q|$ between the average charge inside the interval and the charge label  of the enhanced SRE component. Based on the analytical argument below, we expect that all R\'enyi entropies show an analogous behavior.

The aforementioned behavior originates in the simple manner in which the two-defect zero mode $|\Psi_p\rangle$ modifies the spectrum of the correlation matrix. Letting $\mathcal{P}_A$ denote the projection operator onto subsystem $A$, the correlation matrix of Eq.~\eqref{CijSumk} becomes
\begin{equation}
\begin{split}
C^{\left(p\right)} & = \mathcal{P}_A\left[\sum_k |\phi_k\rangle\langle\phi_k| + |\Psi_p\rangle\langle\Psi_p|\right]\mathcal{P}_A \\
& = C^{\left(p=1\right)} + \left(1-p\right)|\Psi_1\rangle\langle\Psi_1|,
\end{split}
\end{equation}
where $|\phi_k\rangle$ are the extended lower-band single-particle eigenstates that are all fully occupied at the ground state, and where we used the fact that $\mathcal{P}_A|\Psi_1\rangle = |\Psi_1\rangle$ and $\mathcal{P}_A|\Psi_2\rangle=0$, up to exponentially small corrections. This latter property of the localized zero modes, along with the fact that they are orthogonal to all extended eigenstates $|\phi_k\rangle$, also implies that $\left[C^{\left(p=1\right)},|\Psi_1\rangle\langle\Psi_1|\right]=0$ and $C^{\left(p\right)}|\Psi_1\rangle=\left(1-p\right)|\Psi_1\rangle$.

In total, we conclude that the effect of the two-defect 
localized mode amounts to the addition of an independent level, with eigenvalue $1-p$, to the spectrum of the correlation matrix, while all of its other eigenvalues do not change as the parameter $p$ is tuned. This is confirmed by examining the bottom row of Fig.~\ref{figSREPBCkink1s3sSublattice}, where we plot the correlation matrix spectrum for different values of $p$. For convenience, we introduced the function $-\log\sin\left(\pi\lambda_i\right)$, which resolves the eigenvalues $\lambda_i$ close to $1$ and $0$, allowing us to better visualize the spectrum.

\begin{figure}[t]
\centering
\includegraphics[width=\columnwidth]{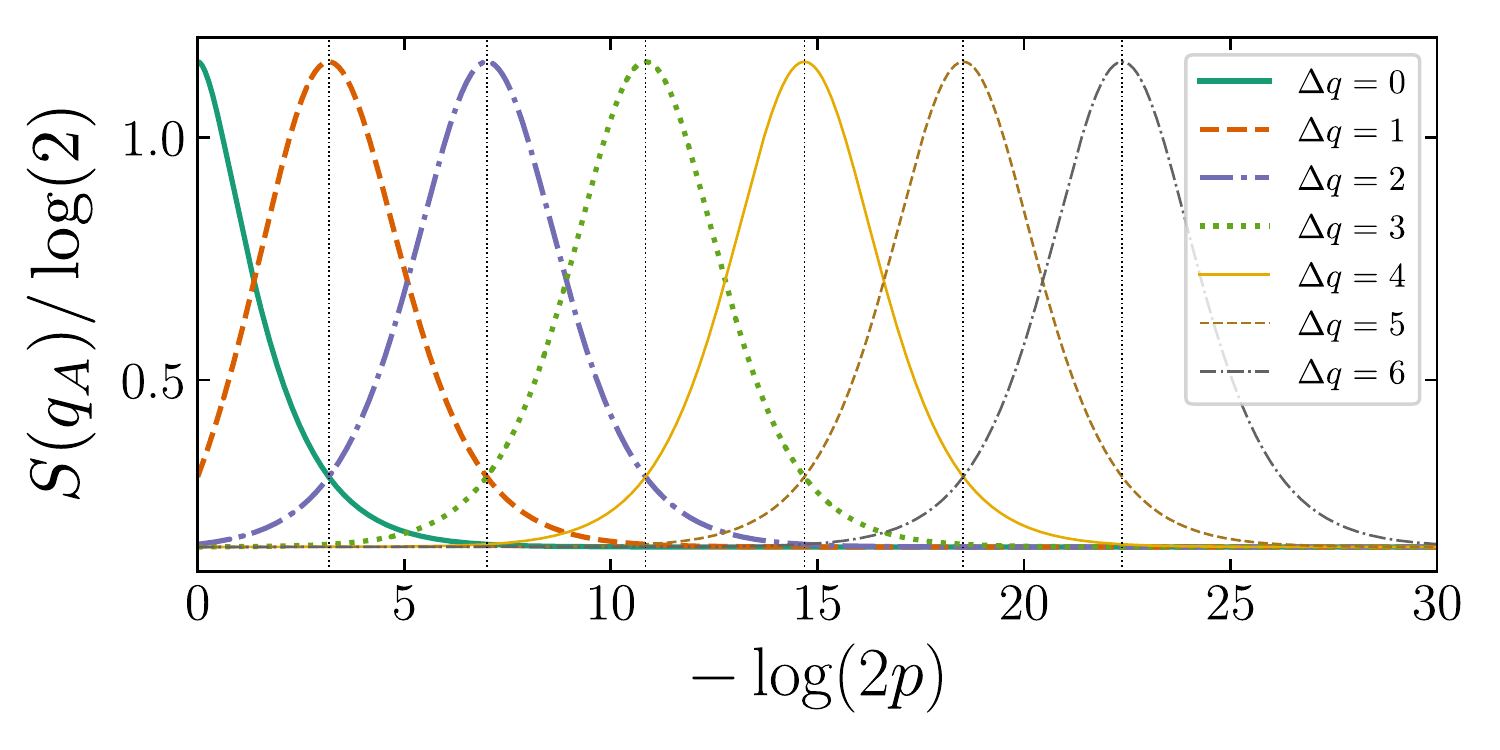}
\caption{\label{figExactSREZM} Asymptotically exact predictions for the SRE of the first few charge sectors for an interval containing a defect, and in the presence of an excited asymmetrically localized zero mode, based on Eq.~\eqref{AnalyticalvNSRE_Def}.
Different colors correspond to different charge sectors; vertical dotted lines indicate the values of $p$ where the level crossings occur, which indeed correspond to the maxima of the SREs.}
\end{figure}

Equivalently, the single-particle spectrum of the entanglement Hamiltonian features an additional pseudo-energy level $\epsilon_{\rm{zero}}=\log\!\left(p/\!\left(1-p\right)\right)$, which is determined by the value of $p$, while all the other pseudo-energies are independent of it. According to Eq.~\eqref{ChargedMomentFromSpectrum}, this modifies the charged moments of the subsystem into
\begin{equation}
\mathcal{Z}_n^{\left(p\right)}\!\left(\alpha\right)=\left[p^n+\left(1-p\right)^n e^{i\alpha}\right]\mathcal{Z}_n^{\rm{def}}\!\left(\alpha\right),
\end{equation}
where $\mathcal{Z}_n^{\rm{def}}\!\left(\alpha\right)$ is the exact result reported in Eq.~\eqref{ExactChargedMoments}. The effect on the SRPFs can then be deduced from Eq.~\eqref{eq:CalU1}, yielding
\begin{equation}
\mathcal{Z}_n^{\left(p\right)}\!\left(q_A\right)=p^n\mathcal{Z}_n^{\rm{def}}\!\left(q_A\right)+\left(1-p\right)^n\mathcal{Z}_n^{\rm{def}}\!\left(q_A-1\right).
\label{ExactChragedPartitionFunctionZM}
\end{equation}
Note that this indeed reduces to the result in Eq.~\eqref{2DefectFullyDimerizedSRPF} derived in the fully dimerized case. Plugging in the result of Eq.~\eqref{AnalyticalSRPF_Def}, we find that the SREs are given by
\begin{equation}
\begin{split}
    S_n^{\left(p\right)}\!\left(q_A\right)&=S_n^{\text{def}}\!\left(q_A\right)+\frac{1}{1-n}\log\!\left[\frac{p^n+\left(1-p\right)^n e^{n\epsilon\Delta q}}{\left(p+\left(1-p\right)e^{\epsilon\Delta q}\right)^n}\right]\\
    &=S_n^{\text{def}}\!\left(q_A\right)+\frac{1}{1-n}\log\!\left[\frac{1+e^{n\left(\epsilon_{\text{zero}}-\epsilon\Delta q \right)}}{\left(1+e^{\left(\epsilon_{\text{zero}}-\epsilon\Delta q \right)}\right)^n}\right],
\end{split}
\label{AnalyticalRenyiSRE_Def}
\end{equation}
and, taking the $n\rightarrow 1$ limit,
\begin{equation}
\begin{split}
    S^{\left(p\right)}\!\left(q_A\right)=&\,\,S^{\text{def}}\!\left(q_A\right)-\frac{1}{1+e^{\epsilon_{\text{zero}}-\epsilon\Delta q}}\log\!\left(\frac{1}{1+e^{\epsilon_{\text{zero}}-\epsilon\Delta q}}\right)\\
    &-\frac{e^{\epsilon_{\text{zero}}-\epsilon\Delta q}}{1+e^{\epsilon_{\text{zero}}-\epsilon\Delta q}}\log\!\left(\frac{e^{\epsilon_{\text{zero}}-\epsilon\Delta q}}{1+e^{\epsilon_{\text{zero}}-\epsilon\Delta q }}\right).
\end{split}
\label{AnalyticalvNSRE_Def}
\end{equation}
We remind the reader that the SREs $S_n^{\text{def}}\!\left(q_A\right)$ and $S^{\text{def}}\!\left(q_A\right)$ are, in fact, independent of $q_A$. Equations ~\eqref{AnalyticalRenyiSRE_Def} and \eqref{AnalyticalvNSRE_Def} show that, for any $n$ and within any charge sector, the two-defect zero mode can lead to a maximal excess entropy of $\log 2$. For a certain value of $p$, the SRE  $S\!\left(q_A\right)$ in a certain charge sector $q_A$ can thus reach a maximal value that exceeds the SREs of the other sectors. Moreover, for any choice of $q_A$, this maximal breaking of entanglement equipartition occurs exactly for the value of $p$ such that $\epsilon_{\text{zero}}=\epsilon\Delta q$, that is, when $\epsilon_{\rm{zero}}$ is degenerate with another ($p$-independent) single-particle level of the entanglement Hamiltonian. This behavior is illustrated in Fig.~\ref{figExactSREZM}, and is confirmed numerically by the results presented in Fig.~\ref{figSREPBCkink1s3sSublattice}.

This phenomenon may again be intuitively understood through the statistical mechanics perspective introduced in Sec.~\ref{AwayFromDimerized_StatMech}. There, we showed how the addition of an average charge constraint to the RDM can turn the SRE of any charge sector into the dominant contribution to the total entropy of the constrained RDM (recall Eq.~\eqref{ConstrainedEntropyDecomposition}). This provides a way to analyze the SRE in terms of the single-particle entanglement spectrum $\left\{\epsilon_i\right\}$ and of a fictitious chemical potential $\mu$ set by the constraint.

In the presence of the excited zero mode, the spectrum ${\epsilon_i}$ is comprised of the pseudo-energies
\begin{equation}
\epsilon_{l}=\epsilon l-\epsilon\ell,\qquad 1\le l\le2\ell-1,
\label{BulkEntanglementSpectrum}
\end{equation}
which constitute the "bulk" spectrum, along with the single independent level $\epsilon_{\rm{zero}}$. The "bulk" spectrum in Eq.~\eqref{BulkEntanglementSpectrum} was inferred from the (in principle infinite) spectrum of Eq.~\eqref{ModHEigvalsDef}, by taking into account the appropriate dimension of the entanglement Hamiltonian (which should correspond to $2\ell$ single-particle modes in total). It is therefore important to note that Eq.~\eqref{BulkEntanglementSpectrum} correctly captures the pseudo-energies away from the edges of the spectrum, where deviations are expected to be observed (see Ref.~\cite{Eisler_2020} and Fig.~\ref{figSREPBCkink1s3sSublattice}).
As $p$ can be modified freely in the ground state, we may think of $\epsilon_{\rm{zero}}$ as a free parameter that can be varied between $-\infty$ (corresponding to $p=1$) and $\infty$ (corresponding to $p=0$). Hence, given a specific choice of an average charge $q_A$, the choice of $\epsilon_{\rm{zero}}$ will obviously determine the value of $\mu$through Eq.~\eqref{ChargeConstraint}, as the only free parameter left.

In Appendix~\ref{ChemicalPotentialAppendix} we show that, for any choice of $\epsilon_{\rm{zero}}$, we have $\epsilon_{q_A-1}<\mu<\epsilon_{q_A+1}$, 
and that a significant entropy contribution (according to Eq.~\eqref{EntropyWithChemicalPotential}) comes from the levels $\epsilon_{\rm{zero}}$ and $\epsilon_{q_A}$ if $\epsilon_{\rm{zero}}\approx\epsilon_{q_A}\approx\mu$. Looking in the vicinity for a value of $\epsilon_{\rm{zero}}$ that maximizes the SRE $S\!\left(q_A\right)$, we then find that such a maximum is reached when $\epsilon_{\rm{zero}}=\epsilon_{q_A}=\epsilon\Delta q$, showing the correspondence between the entanglement spectrum degeneracy and the breaking of entanglement equipartition.

It should be noted that this argument holds only when $\epsilon$ is finite, and it therefore applies only away from the fully dimerized limit. In particular, when $\delta\rightarrow 1$, the "bulk" eigenvalues of the RDM approach either 0 or 1, and the corresponding pseudo-energies approach $\pm\infty$, which are to be compared against $\epsilon_{\rm{zero}}$. This means, that as $\delta$ approaches $1$, the maxima in the charge sectors will appear at smaller (or closer to $1$) values of $p$, and when $\delta=1$ only one maximum corresponding the $q_A=\ell$ sector can appear, namely for $p=1/2$.\\
We also reiterate the fact that Eq.~\eqref{BulkEntanglementSpectrum} fails to capture the edges of the entanglement spectrum in a finite interval. Our analysis of the equipartition breaking phenomenon should be therefore seen as restricted to values of $q_A$ that are not near the extremal possible values of the charge, namely $q_A=0,2\ell$.

Finally, we find it important to stress that the enhancement for sectors with far-from-the-average charges are only present in the SREs, and not in the SRPFs. In particular, the probabilities for the charge fluctuations remain suppressed when the deviation from the average charge value is large. This guarantees that the total, fluctuational and configurational entropies do not change when $p$ is very close to $0$ or $1$, even if a certain SRE component is enhanced. \\

To summarize, in this section we treated the possibility of the zero modes, which are localized at the topological defects, being excited, and studied the effects of such excitations on the SREs. We focused our analysis on an excited zero mode that is supported on two defects, of which the subsystem contains only a single defect (see Eq.~\eqref{EpsPsi}, after which we also briefly comment on the realization of such a state). As the main results of this section, Eqs.~\eqref{AnalyticalRenyiSRE_Def} and \eqref{AnalyticalvNSRE_Def} show that in this case entanglement equipartition is broken by an enhanced entropy of a particular charge sector. The charge label of this sector is not necessarily close to the average subsystem charge, and is determined by the relative probability weight of the zero mode on the two defects. We observed that the enhanced SRE reaches a maximum exactly when the relative weight is such that a degeneracy occurs in the entanglement spectrum, thus relating these results to the findings of Sec.~\ref{AwayFromDimerized}.

\section{Conclusions and outlook}\label{sec:conclusions}
Topological defects like solitons appear in many different areas of physics, from gauge theories~\cite{rajaraman1982solitons} and polymers~\cite{heeger1988solitons} to solid state physics~\cite{peierls} and quantum simulation~\cite{roy2023soliton,rylands2020photon,Horvath_2022SG}.  In this work we have investigated such defects through the lens of symmetry-resolved entanglement measures using the paradigmatic example of the SSH model.  We have seen,  using a  combination of analytical techniques backed up by numerical computations, that they support a hidden entanglement structure which is unveiled by considering their interplay with charge.  In particular, while the total entropy only provides a mundane viewpoint on the physics of the defects, we observe a redistribution of the entanglement entropy between its configurational and fluctuational components due to the presence of the defect.

The symmetry-resolved entropies also unveil an unexpected connection between the entanglement properties of the trivial and topological phases of the SSH chain. Namely, in both phases all charge sectors of the same parity have an equal entropy contribution, but the contributions of sectors of the opposite parity differ (note that a similar behavior was observed in the gapped XXZ chain for a subsystem with an odd number of sites \cite{CalabreseFCS_2020}). Thus, the conventional entanglement equipartition is broken into an alternating pattern of two possible values of the symmetry-resolved entropy. Moreover, exactly the same two values appear in both phases, only the alternating pattern is flipped. Intriguingly, the equipartition of entanglement is recovered only when the defect is included in the subsystem.

Furthermore, 
upon exciting the zero modes which are localized on various defects, we can find a significant enhancement of the entanglement also in charge sectors which are far from the average charge value. That is, the systematic breaking of entanglement equipartition can be observed for a subsystem with a defect and it is triggered by the asymmetric localization of zero modes on topological defects. Keeping in mind the fragile nature of these excitations, a simple physical setup is also proposed in Sec. \ref{SectionWithZM} in which such asymmetrically localized modes can be present (see the discussion following Eq.~\eqref{EpsPsi}).

Through a general argument inspired by statistical mechanics, we were able to relate these instances of equipartition breakdown to degeneracies in the spectrum of the entanglement Hamiltonian. More concretely, the entropy of a certain charge sector is enhanced when a degeneracy appears in the single-particle pseudo-energy that matches the charge value within that sector (when counting the pseudo-energies from the bottom of the entanglement spectrum).
This suggests a mechanism that could lead to similar phenomena occurring in a wider class of physical models.

For convenience, throughout our calculations we assumed PBC and investigated the case of only two defects. Nevertheless, our results straightforwardly generalize to other setups, including chains with more than two defects and finite-size chains with edge modes at the boundaries. In particular, in the latter case and with no excited zero modes, it is easy to see that if the subsystem contains one edge of the chain (e.g., $A=[1,\ell]$), entanglement equipartition holds in both the trivial phase and the topological phase. Furthermore, the values of the SREs are independent both of whether the subsystem boundary that is in the interior of the system cuts a strong or a weak bond, and of whether or not the subsystem contains defects.
This is due to the fact that the entanglement spectrum is determined only by the interior subsystem boundary (i.e., it is the spectrum of a half-infinite subsystem), and this spectrum is in any case non-degenerate and equidistant. In the topological phase when zero energy edge modes are excited, the mechanism of eigenvalue-crossing can lead to the same enhancement in the SREs as the one observed for excited zero modes localized at defects, with the same functional expression for the excess entropy.

Our findings are based on the analysis of a non-interacting model; however, both the SSH model and its continuum counterpart can be viewed as describing the fermionic fluctuations about topological excitations in interacting systems. In these systems, the solitons are dynamical, semiclassical solutions of the many-body problem, and our results can therefore provide the leading contribution to their entanglement too [cf.~calculations of the entanglement in the theory of quantum generalized hydrodynamics (see, e.g.,~\cite{ruggiero2020quantum,scopa2021exact,ruggiero2021quantum,collura2020domain})]. 

One could also consider topological defects in genuinely interacting models.  
Obvious candidates for this study are carbon nanotubes \cite{Moca_2020} or the spin-1 Heisenberg chain in the presence of topological defects.
In this case, one can first investigate the system in the AKLT limit \cite{PhysRevLett.59.799} and consider a defect which causes the system to be in the AKLT ground state to one side, and a ferromagnetic ground state (or any other product state) to the other.  We report preliminary results on this in Appendix~\ref{app:AKTL}. Our calculation shows a similar structure of symmetry-resolved entanglement (where the symmetry considered is the conservation of the spin $z$-component) compared to what we found for the fully dimerized SSH chain. The investigation of these models is ongoing, and further results will be deferred to subsequent publications.

\begin{acknowledgments}
The authors are grateful to Filiberto Ares for taking part in numerous discussions, sharing ideas and for giving useful pieces of advice. We are indebted to János Asbóth for discussions, for his feedback on the manuscript and for proposing a physically motivated setup to prepare asymmetrically localized zero modes. We also thank Gilles Parez, Giuseppe Di Giulio, Eran Sela, Moshe Goldstein, and Pasquale Calabrese for useful discussions and feedback. S.F.~is supported by the Israel Science Foundation (ISF) and the Directorate for Defense Research and Development (DDR\&D) grant No.~3427/21, by the US-Israel Binational Science Foundation (BSF) Grant No.~2020072, and by the Azrieli Foundation Fellows program. D.X.H., C.R., and S.S.~acknowledge support from ERC under Consolidator grant number 771536 (NEMO).
\end{acknowledgments}

\bibliography{SymmetryResolvedSSH}

\newpage

\appendix

\section{Fully dimerized limit of the SSH chain}\label{AppendixCorrelMatrices}
In the fully dimerized limit of the SSH chain, the correlation matrix of a single interval of length $\ell$ is a $2\ell\times2\ell$ matrix which can be trivially written in terms of block matrices. We refer to the three relevant cases "trivial," "topological," and "defect" by denoting the corresponding matrices as $C_\text{triv}$, $C_\text{top}$ and $C_\text{def}$. In particular, we have
\begin{equation}
\begin{split} 
& C_\text{triv}=
\left(\begin{array}{ccccc}
(1/2)_{2\times2} & (0)_{2\times2} & \cdots &  & (0)_{2\times2}\\
(0)_{2\times2} & (1/2)_{2\times2} &  &  & \vdots\\
\vdots &  & \ddots\\
 &  &  &  & (0)_{2\times2}\\
(0)_{2\times2} & \cdots &  & (0)_{2\times2} & (1/2)_{2\times2}
\end{array}\right)
\end{split}
\end{equation}
in the trivial phase and

\begin{equation}
\begin{split} 
 & C_\text{top}=
 \left(\begin{array}{cccccc}
(1/2)_{1\times1} & (0)_{1\times2}  & \cdots &  &  & (0)_{1\times2}\\
(0)_{2\times1}   &  &  &  &  &   \\
 \vdots     & \block(4,4){C_\text{triv}} & \vdots  \\
&  &  &  & &\\
 &  &  &  &  & \\
 &  &  &  &   &(0)_{1\times2}\\
 (0)_{2\times1}& \cdots &  &  & (0)_{1\times2} & (1/2)_{1\times1} 
\end{array}\right)
\end{split}
\end{equation}
in the topological phase. If the interval includes a defect (with topological phase cells on its left and trivial phase cells on its right) and we assume that the corresponding zero mode is not excited, then we have

\begin{equation}
\begin{split}  
 & C_{\text{def}}=\left(\begin{array}{ccccccc}
(1/2)_{2\times2} & (0)_{2\times2} & \cdots &  &  & (0)_{2\times2} & (0)_{1\times2}\\
(0)_{2\times2} & \ddots &  &  &  & \vdots & \vdots\\
\vdots & &\block(3,3){\tilde{C}_\text{def}} &  & \\
 &  &  & & & \\
 &  & & & & & \\
(0)_{2\times2} & \cdots &  &  &  & \ddots & (0)_{1\times2}\\
(0)_{2\times1} & \cdots &  &  &  & (0)_{2\times1} & (1/2)_{1\times1}
\end{array}\right)
\end{split}
\end{equation}
with
\begin{equation}
\begin{split}  
 & \tilde{C}_{\text{def}}=\left(\begin{array}{ccc}
(1/2)_{2\times2} & (0)_{2\times.} & (0)_{2\times2}\\
(0)_{.\times2} & (d)_{.\times.} & (0)_{.\times2}\\
(0)_{2\times2} & (0)_{2\times.} & (1/2)_{2\times2}\\
\end{array}\right)\,,
\end{split}
\end{equation}
and depending on the type of the defect, 1s or 3s, the matrix $(d)_{.\times.}$ is either a $1\times1$ matrix $(d)_{1\times1}=0$ or a $3\times3$ matrix 
\begin{equation}
(d)_{3\times3}=\left(\begin{array}{ccc}
\frac{1}{4} & -\frac{1}{2\sqrt{2}} & \frac{1}{4}\\
-\frac{1}{2\sqrt{2}} & \frac{1}{2} & -\frac{1}{2\sqrt{2}}\\
\frac{1}{4} & -\frac{1}{2\sqrt{2}} & \frac{1}{4}
\end{array}\right)\,,
\end{equation}
with eigenvalues $1,0,0$, respectively. The dots in the block
matrices indicate the consistency with the right size referring to either 1 or 3.

\section{Derivation of the main analytical expressions}
\subsection{Charged moments}
\label{AnalyticalChargedMomentsAppendix}
In this appendix we provide a detailed derivation of our analytical formulas for the boundary contributions to the charged moments, reported in Section \ref{AwayFromDimerized}. From the general expression in Eq.~\eqref{ChargedMomentFromSpectrum} for charged moments in terms of the single-particle entanglement spectrum, along with the result in Eq.~\eqref{ModHEigvals1} for this spectrum in the case of a single cut at a strong bond, we observe that
\begin{equation}
\begin{split}
Z^{\rm{strong}}_{n}\!\left(\alpha\right)=&\,\frac{1+e^{i\alpha}}{2^{n}}\\
&\times\prod_{l\ge1}\!\left[\frac{e^{i\alpha}\!\left(1+e^{-2n\epsilon l+i\alpha}\right)\!\left(1+e^{-2n\epsilon l-i\alpha}\right)}{\left(1+e^{-2\epsilon l}\right)^{2n}}\right]\!.
\end{split}
\end{equation}
The usual procedure for $\alpha=0$ (i.e., the one employed when the unresolved entropy is the quantity of interest) dictates that we should multiply the terms in the product up to $l=\infty$, as these terms quickly converge to $1$. In the case where $\alpha\neq0$, we must take out the additional $e^{i\alpha}$ factor out of the product before replacing it with an infinite product. This results in an overall phase factor that counts the number of entanglement Hamiltonian modes with negative pseudo-energies. This number is infinite when considering the problem of a half-infinite subsystem, making the overall phase factor ill-defined; we therefore regard it as a proportionality factor that can be exactly specified only for the finite subsystem, i.e., only after the two boundary contributions are combined. We discuss this issue in Section \ref{AwayFromDimerized}.

With that in mind, we write
\begin{equation}
Z^{\rm{strong}}_{n}\!\left(\alpha\right) \propto\frac{1+e^{i\alpha}}{2^{n}}\prod_{l=1}^{\infty}\left[\frac{\left|1+e^{-2n\epsilon l+i\alpha}\right|^2}{\left(1+e^{-2\epsilon l}\right)^{2n}}\right].
\end{equation}
To move forward, we introduce the definitions of the well-studied Jacobi theta functions \cite{whittaker_watson_1996}
\begin{equation}
\begin{split}
\theta_{2}\!\left(\omega|\zeta\right)= &\sum_{m=-\infty}^{\infty}e^{i\left(2m+1\right)\omega}\zeta^{\left(m+\frac{1}{2}\right)^2}= 2\zeta^{1/4}\cos\omega \\
& \times\prod_{l=1}^{\infty}\left[\left(1-\zeta^{2l}\right)\left(1+2\cos\left(2\omega\right)\zeta^{2l}+\zeta^{4l}\right)\right], \\
\theta_{3}\!\left(\omega|\zeta\right)=&\sum_{m=-\infty}^{\infty}e^{2im\omega}\zeta^{m^2}\\
=&\,\prod_{l=1}^{\infty}\left[\left(1-\zeta^{2l}\right)\left(1+2\cos\left(2\omega\right)\zeta^{2l-1}+\zeta^{4l-2}\right)\right],
\end{split}
\label{ThetaFunctionsDefinition}
\end{equation}
and recognize that
\begin{equation}
Z^{\rm{strong}}_{n}\!\left(\alpha\right)\propto\frac{\theta_{2}\!\left(\frac{\alpha}{2}|e^{-n\epsilon}\right)}{\theta_{2}\!\left(e^{-\epsilon}\right)^{n}}\prod_{l=1}^{\infty}\frac{\left(1-e^{-2\epsilon l}\right)^{n}}{\left(1-e^{-2n\epsilon l}\right)},
\end{equation}
where we again employed the notation $\theta_{j}\!\left(\zeta\right)=\theta_{j}\!\left(0|\zeta\right)$.

Now, recalling the definitions of $k_{n}$ and $k_{n}'=\sqrt{1-{k_{n}}^{2}}$ through
the relation
\begin{equation}
n\epsilon=\pi\frac{I\!\left(k_{n}'\right)}{I\!\left(k_{n}\right)},
\end{equation}
and using the identity \cite{whittaker_watson_1996}
\begin{equation}
\prod_{l=1}^{\infty}\left(1-\zeta^{2l}\right)=\left[\frac{\tilde{k}\sqrt{1-\tilde{k}^{2}}}{4\zeta^{1/2}}\right]^{1/6}\theta_{3}\!\left(\zeta\right),
\label{InfiniteProdIdentity}
\end{equation}
which applies to $\zeta=\exp\!\left[-\pi I\!\left(\sqrt{1-\tilde{k}^{2}}\right)/I\!\left(\tilde{k}\right)\right]$, we  obtain
\begin{equation}
Z^{\rm{strong}}_{n}\!\left(\alpha\right)\propto\frac{\theta_{2}\!\left(\frac{\alpha}{2}|e^{-n\epsilon}\right)}{2^{\left(n-1\right)/3}\theta_{2}\!\left(e^{-\epsilon}\right)^{n}}\!\left(\frac{k^{n}{k'}^{n}}{k_{n}k_{n}'}\right)^{1/6}\frac{\theta_{3}\!\left(e^{-\epsilon}\right)^{n}}{\theta_{3}\!\left(e^{-n\epsilon}\right)}.
\end{equation}
By using the identity $k=\left[\theta_{2}\!\left(e^{-\epsilon}\right)/\theta_{3}\!\left(e^{-\epsilon}\right)\right]^{2}$ \cite{whittaker_watson_1996},
we may finally write
\begin{equation}
Z^\text{strong}_{n}\!\left(\alpha\right)\propto\frac{1}{2^{\left(n-1\right)/3}}\left(\frac{{k'}^{n}}{k_{n}k_{n}'k^{2n}}\right)^{1/6}\frac{\theta_{2}\!\left(\frac{\alpha}{2}|e^{-n\epsilon}\right)}{\theta_{3}\!\left(e^{-n\epsilon}\right)}.
\end{equation}

A similar analysis leads to the result for the charged moments in the case where the single cut is located at a weak bond. In that case, the single-particle entanglement spectrum from Eq.~\eqref{ModHEigvals2} leads to the expression
\begin{equation}
Z^{\rm{weak}}_{n}\!\left(\alpha\right)=\prod_{l\ge1}\left[\frac{e^{i\alpha}\left|1+e^{-n\epsilon\left(2l-1\right)+i\alpha}\right|^2}{\left(1+e^{-\epsilon\left(2l-1\right)}\right)^{2n}}\right],
\end{equation}
which by the same argument as before can be written as
\begin{align}
Z^{\rm{weak}}_{n}\!\left(\alpha\right) & \propto\prod_{l=1}^{\infty}\left[\frac{\left|1+e^{-n\epsilon \left(2l-1\right)+i\alpha}\right|^2}{\left(1+e^{-\epsilon\left(2l-1\right)}\right)^{2n}}\right]\nonumber \\
 & =\frac{\theta_{3}\!\left(\frac{\alpha}{2}|e^{-n\epsilon}\right)}{\theta_{3}\!\left(e^{-\epsilon}\right)^{n}}\prod_{l=1}^{\infty}\frac{\left(1-e^{-2\epsilon l}\right)^{n}}{\left(1-e^{-2n\epsilon l}\right)}.
\end{align}
Applying the identity \eqref{InfiniteProdIdentity} as before then yields
\begin{equation}
Z^\text{weak}_{n}\!\left(\alpha\right)\propto\frac{1}{2^{\left(n-1\right)/3}}\left(\frac{k^{n}{k'}^{n}}{k_{n}k_{n}'}\right)^{1/6}\frac{\theta_{3}\!\left(\frac{\alpha}{2}|e^{-n\epsilon}\right)}{\theta_{3}\!\left(e^{-n\epsilon}\right)}.
\end{equation}

\subsection{Charge resolved partition functions}
\label{AnalyticalSRPFAppendix}
In this appendix we explain in detail how the Fourier transform of Eq.~\eqref{eq:CalU1}, which is the crucial component in the computation of the SRPFs once the charged moments have been derived, can be done analytically in the case of the charged moments in Eq.~\eqref{ExactChargedMoments}.

We first observe, based on the definition of the theta functions in Eq.~\eqref{ThetaFunctionsDefinition}, that
\begin{equation}
\left[\theta_{2}\!\left(\omega|\zeta\right)\right]^2=\sum_{m,l=-\infty}^{\infty}\!e^{2i\omega\left(m+l+1\right)}\zeta^{m^2+l^2+m+l+\frac{1}{2}}.
\end{equation}
We then replace the summation indices $m,l$ with $r,l$, defining $r=m+l+1$. This allows us to write
\begin{equation}
\left[\theta_{2}\!\left(\omega|\zeta\right)\right]^2=\zeta^{-\frac{1}{2}}\sum_{r,l=-\infty}^{\infty}\!e^{2i\omega r}\zeta^{\left(r-l\right)^2+\left(l+1\right)^2-r}.
\end{equation}
Next, we identify $e^{i\alpha\ell}\left[\theta_{2}\!\left(\frac{\alpha}{2}|e^{-n\epsilon}\right)\right]^2$ as the $\alpha$-dependent factor in the expression for $Z_n^{\text{top}}\!\left(\alpha\right)$ in Eq.~\eqref{ExactChargedMoments}, which will thus be the only factor modified under the Fourier transform of Eq.~\eqref{eq:CalU1}. Writing $q_A=\ell+\Delta q$, we see that
\begin{equation}
\begin{split}
&\int_{-\pi}^{\pi}\frac{\mathrm{d}\alpha}{2\pi}e^{i\alpha\ell}\left[\theta_{2}\!\left(\frac{\alpha}{2}{\bigg |}e^{-n\epsilon}\right)\right]^2e^{-i\alpha q_{A}} \\
&=e^{n\epsilon/2}\sum_{l=-\infty}^{\infty}\!e^{-n\epsilon\left[\left(\Delta q-l\right)^2+\left(l+1\right)^2-\Delta q\right]} \\
&=e^{-\frac{1}{2}n\epsilon\,\left(\Delta q\right)^2}\sum_{l=-\infty}^{\infty}\!\exp\!\left[-2n\epsilon\left(l-\frac{\Delta q - 1}{2}\right)^2\right].
\end{split}
\end{equation}
Given that the index $l$ is summed over all integers, and that $\Delta q$ is itself an integer, we conclude that the resultant value of the sum depends only on the parity of $\Delta q$. We may further determine the exact value of the sum by examining Eq.~\eqref{ThetaFunctionsDefinition}, obtaining
\begin{equation}
\begin{split}
&\int_{-\pi}^{\pi}\frac{\mathrm{d}\alpha}{2\pi}e^{i\alpha\ell}\left[\theta_{2}\!\left(\frac{\alpha}{2}{\bigg |}e^{-n\epsilon}\right)\right]^2e^{-i\alpha q_{A}}\\
&=e^{-\frac{1}{2}n\epsilon\,\left(\Delta q\right)^2}\begin{cases}
\theta_3\!\left(e^{-2n\epsilon}\right) & \Delta q {\text{ is odd}} \\
\theta_2\!\left(e^{-2n\epsilon}\right) & \Delta q {\text{ is even}}\,.
\end{cases}
\end{split}
\label{Theta2Fourier}
\end{equation}
This then leads to the result for the SRPF in the topological phase, reported in Eq.~\eqref{AnalyticalSRPF_TopTriv}.

A procedure similar to the one that led to Eq.~\eqref{Theta2Fourier} can be applied when considering the Fourier transform of the two other charged moments appearing in Eq.~\eqref{ExactChargedMoments}. For the $\alpha$-dependent factor in the expression for $Z_n^{\text{triv}}\!\left(\alpha\right)$, we find that its Fourier transform is given by
\begin{equation}
\begin{split}
&\int_{-\pi}^{\pi}\frac{\mathrm{d}\alpha}{2\pi}e^{i\alpha\ell}\left[\theta_{3}\!\left(\frac{\alpha}{2}{\bigg |}e^{-n\epsilon}\right)\right]^2e^{-i\alpha q_{A}}\\
&=e^{-\frac{1}{2}n\epsilon\,\left(\Delta q\right)^2}\begin{cases}
\theta_2\!\left(e^{-2n\epsilon}\right) & \Delta q {\text{ is odd}} \\
\theta_3\!\left(e^{-2n\epsilon}\right) & \Delta q {\text{ is even}}\,,
\end{cases}
\end{split}
\end{equation}
and for the corresponding factor in the expression for $Z_n^{\text{def}}\!\left(\alpha\right)$, the same procedure yields
\begin{equation}
\begin{split}
&\int_{-\pi}^{\pi}\frac{\mathrm{d}\alpha}{2\pi}e^{i\alpha\left(\ell-1/2\right)}\theta_3\!\left(\frac{\alpha}{2}{\bigg |}e^{-n\epsilon}\right)\theta_2\!\left(\frac{\alpha}{2}{\bigg |}e^{-n\epsilon}\right)e^{-i\alpha q_{A}}\\
&=\frac{1}{2}e^{-\frac{1}{2}n\epsilon\,\left(\Delta q+\frac{1}{2}\right)^2}\theta_2\!\left(e^{-n\epsilon/2}\right).
\end{split}
\end{equation}
This allows to complete the computation of the SRPFs also for the trivial phase and for the case where the interval includes a defect. The results are reported in Eqs.~\eqref{AnalyticalSRPF_TopTriv} and \eqref{AnalyticalSRPF_Def}.

\section{Charge-resolved entropy with a zero mode contribution}
\label{ChemicalPotentialAppendix}
In this appendix we provide technical details of the argument made in Section \ref{SectionWithZM} regarding the correspondence between a degeneracy of the single-particle entanglement spectrum (due to a contribution from the zero mode supported on two defects, one of which is inside the subsystem of interest) and the breaking of entanglement equipartition.  For this purpose, 
we analyze the relation in Eq.~\eqref{ConstrainedEntropyDecomposition} between the SREs and a charge distribution that is modified due to an average charge constraint. The constraint (see Eq.~\eqref{ChargeConstraint}) shifts the single-particle entanglement spectrum (given by Eq.~\eqref{BulkEntanglementSpectrum} along with $\epsilon_{\text{zero}}$) by a fictitious chemical potential $\mu$.

Consider the charge sector with charge $q_A$. To probe the SRE of this sector, we impose the average charge constraint of Eq.~\eqref{ChargeConstraint}, yielding
\begin{equation}
\begin{split}
q_A & =\frac{1}{e^{\epsilon_{{\rm zero}}-\mu}+1}+\sum_{l=1}^{2\ell-1}\frac{1}{e^{\epsilon_{l}-\mu}+1} \\
& =\frac{1}{e^{\epsilon_{{\rm zero}}-\mu}+1}+\sum_{l=0}^{2\ell-2}\frac{1}{e^{-\left(\mu+\left(\ell-1\right)\epsilon\right)}e^{\epsilon l}+1}\,.
\end{split}
\label{ChargeConstraint_ZeroMode}
\end{equation}
As we claimed in Sec.~\ref{SectionWithZM}, we can view $\epsilon_{{\rm zero}}$ as a free parameter ranging between $\pm\infty$, which then determines $\mu$ through Eq.~\eqref{ChargeConstraint_ZeroMode}, given that all other spectrum levels are fixed. For $\ell\gg1$ we may write 
\begin{equation}
q_A=\frac{1}{e^{\epsilon_{{\rm zero}}-\mu}+1}+\int_{0}^{\infty}\frac{dx}{e^{-\left(\mu+\left(\ell-1\right)\epsilon\right)}e^{\epsilon x}+1}+\Delta_{\epsilon,\mu}\,,
\end{equation}
where $0<\Delta_{\epsilon,\mu}<1$ is a remainder due to the replacement
of the sum with an integral (the sum constitutes a discretized version
of the integral such that at each interval of length $1$ we sample
the integrand at the left edge of the interval, which is the maximum
point of the integrand within the interval). This then leads to the approximation
\begin{equation}
\begin{split}
q_A & =\frac{1}{e^{\epsilon_{{\rm zero}}-\mu}+1}+\frac{1}{\epsilon}\log\!\left(1+e^{\left(\mu+\left(\ell-1\right)\epsilon\right)}\right)+\Delta_{\epsilon,\mu} \\
 & \approx\frac{1}{e^{\epsilon_{{\rm zero}}-\mu}+1}+\frac{\mu}{\epsilon}+\ell-1+\Delta_{\epsilon,\mu}\,,
\end{split}
\end{equation}
or
\begin{equation}
\mu=\epsilon\left[q_A-\frac{1}{e^{\epsilon_{{\rm zero}}-\mu}+1}-\ell+1-\Delta_{\epsilon,\mu}\right].\label{eq:Self-consistent-mu}
\end{equation}
In particular, considering the form of the `bulk' entanglement spectrum in Eq.~\eqref{BulkEntanglementSpectrum}, we have $\epsilon_{q_A-1}<\mu<\epsilon_{q_A+1}$. Therefore, out of the `bulk' spectrum, $\epsilon_{q_A}$ is the only level with which $\mu$ can coincide under the charge constraint we chose. We will now show that the SRE $S\!\left(q_A\right)$ obtains its maximal value (thus leading to a maximal breaking of equipartition) when $\mu$ is very close to $\epsilon_{q_A}$.

Indeed, if $\mu=\epsilon_{q_A}$ then the average occupation of `bulk' levels in Eq.~\eqref{ChargeConstraint_ZeroMode} approximately yields $q_A-1/2$ (here, as in Subsec.~\ref{AwayFromDimerized_StatMech}, we assume that the level $\epsilon_{q_A}$ is far enough from the spectrum edges, and approximate the spectrum as symmetric about this level). To accommodate the constraint in Eq.~\eqref{ChargeConstraint_ZeroMode} we must therefore have $\epsilon_{{\rm zero}}\approx\mu$, meaning that $\epsilon_{{\rm zero}}$ and $\epsilon_{q_A}$ are very close to one another, and also that they both produce the dominant entropy contributions, according to Eq.~\eqref{EntropyWithChemicalPotential}. The entropy contributions from all other levels are much smaller (and vanish in the limit $\delta\to1$) assuming that $\epsilon$ is large enough: the levels $\epsilon_l$ with $l<q_A$ are effectively occupied, and those with $l>q_A$ are effectively empty.

We now look for the value of $\epsilon_{{\rm zero}}$ that maximizes $S\!\left(q_A\right)$, in the vicinity of the value of $\epsilon_{{\rm zero}}$ for which $\mu=\epsilon_{q_A}$. We observe that the charge sector with charge $q_A$ is dominated by two states: the one where $\epsilon_{{\rm zero}}$ is empty and $\epsilon_{q_A}$ is occupied, with its probability proportional to $e^{\epsilon_{{\rm zero}}-\mu}$; and the one where $\epsilon_{{\rm zero}}$ is occupied and $\epsilon_{q_A}$ is empty, with its probability proportional to $e^{\epsilon_{q_A}-\mu}$. It is then straightforward to see that the maximal value of the entropy $S\!\left(q_A\right)$ is reached when the two probabilities are equal, i.e., at the degeneracy point $\epsilon_{{\rm zero}}=\epsilon_{q_A}$.

\section{A spin-1 chain with an AKLT segment and with two defects}\label{app:AKTL}
In this appendix we demonstrate a simple example of a spin-1 system which shows very similar features to the SSH case. In particular, we investigate a system with PBC which is composed of the interacting AKLT chain \cite{PhysRevLett.59.799} on one half, and another spin chain supporting a ground state of a product form on the other half of the system. This way we naturally obtain two defects at the interfaces of the two chains. The ground state of the other spin system can be a ferromagnetic state originating from an interacting spin chain but, for simplicity, we choose a system whose ground state is $\prod_i|0\rangle_i$. 

Of course, extending the system with an essentially independent product state might seem slightly artificial and superfluous. Nevertheless, via this extension we not only naturally obtain two non-trivial interfaces hosting the AKLT spin-1/2 modes but, additionally, we can derive  results for the total and $S^z$-resolved entanglement that are in one-to-one correspondence with the fully dimerized limit of the SSH chain. Alternatively, one could merely consider the AKLT chain with OBC and that way the SSH results for the defect and for topological phases are recovered as well depending on whether the interval contains the edge of the AKLT chain or not and is far away from it.

To be more specific, let us eventually define the model as \begin{equation}
H=\sum_{i=1}^N h_i
\end{equation}
with
\begin{equation}
\begin{split}
h_i&=\overrightarrow{S}_{i}\overrightarrow{S}_{i+1}+\frac{1}{3}\left(\overrightarrow{S}_{i}\overrightarrow{S}_{i+1}\right)^{2} \quad \text{ if } 1\leq i < N/2\\
h_i&=\left(S_i^z\right)^2 \quad\quad\quad\quad\quad\quad\quad\quad\quad\quad \text{ if } N/2< i < N\\
\end{split}
\label{AKLTProductH}
\end{equation}
where $\overrightarrow{S}$ are spin-1 operators, and PBC are imposed via $\overrightarrow{S}_1=\overrightarrow{S}_{N+1}$.
This Hamiltonian commutes with the $J_{\text{tot}}^z$ total spin-$z$ operator, which plays the role of the charge, as well as with its restriction to a subsystem $A$, denoted by $J_A^z$. The ground state of the Hamiltonian is the AKLT ground state with open boundary condition on one side and $\prod_i|0\rangle_i$ on the other side, where $|0\rangle_i$ is the zero spin-$z$ eigenstate of $S_i^z$. 
In this spin system, clearly, the roles of the trivial phase and the topological phase are played by the trivial product state and by the non-trivially entangled AKLT ground state, respectively. This latter state is known to be a valance bond state, which is made up of consecutive spin-singlet states each composed of two spin-1/2 degrees of freedom. These singlets link neighboring lattice sites and two spin-1/2-s  can be regarded as a fractionalization of the original spin-1 degrees of freedom.

Because the ground state of the AKLT with OBC is fourfold degenerate, the total ground state of the Hamiltonian \eqref{AKLTProductH} is fourfold degenerate as well. 
More precisely, the ground states of the spin Hamiltonian \eqref{AKLTProductH} are distinguished by the behavior of the two edge modes of the AKLT part of the chain, which can be regarded as solitary or unpaired spin-1/2-s. These degrees of freedom eventually form a global spin singlet and three global spin-1 triplet states. We therefore label the ground states of the entire chain as $\mid \Updownarrow \rangle_-$ to denote the spin-singlet state (with $J^z_\text{tot}=0$) and as $\mid \Downarrow \rangle$, $\mid \Updownarrow \rangle_+$ and $\mid \Uparrow  \rangle$ to denote the triplets with $J^z_\text{tot}=-1,0,1$, respectively.
This degeneracy corresponds to the fourfold degeneracy of the SSH chain ground state around half-filling, due to the two zero modes localized at its edges.

Finally, we again stress that instead of $h_i=(S_i^z)^2$ we could have chosen
\begin{equation}
h_i=-\left(\overrightarrow{S}_{i}\right)^2-h S_i^z
\end{equation}
still commuting with the the total spin-$z$ operator and yielding a product state as well. In this case the ground states are no longer spin-1 or spin-0 states, but their total spin would be $N/2$ or $N/2+1$ and this would result in a trivial shift for the average value  $\langle J_A^z\rangle$ if the interval includes the defect. 

We now show that, when considering the system described by Eq.~\eqref{AKLTProductH}, the formulas that arise for the SRE are similar to those derived for the SSH chain in its fully dimerized limit.

\subsection{Entropies for $J^z_\text{tot}=\pm1$ states}
We first study the total and SR entropies in the ground states with $J^z_\text{tot}=\pm1$, which correspond to the half-filled ground state of the SSH chain without the zero modes, or involving both of them.
In these cases we can immediately adopt the results of Sec. \ref{FullyDimerizedTotalEntropies} and for brevity, we only present formulas for the entropies such as

\begin{equation}
S_{n}=\begin{cases}
0 & \text{triv}\\
2\log2 & \text{top}\\
\log2 & \text{def}\,,
\end{cases}
\end{equation}
that is, the von Neumann entropy and all the R\'enyi entropies exhibit exactly the same behavior: they are all equal to $0$ in the trivial phase, $2\,\log2$ in the topological phase, and $\log2$ for an interval which contains a defect. 

For both $J^z_{\text{tot}}=\pm1$ states, the SREs can be written as follows:
\begin{equation}
S^{\text{triv}}_{n}\!\left(J^z_A\right)=
0 ~~ \forall J^z_A
\end{equation}
for the trivial phase, 
\begin{equation}
S^{\text{top}}_{n}\!\left(J^z_A\right)=\begin{cases}
\log2 & \text{for } J^z_A=0\\
0 & \text{otherwise}
\end{cases}
\end{equation}
for the topological phase, and finally 
\begin{equation}
S_{n}^{\text{def}}\!\left(J^z_A\right)=
0 ~~ \forall J^z_A.
\end{equation}
As we can see, our considerations made for the dimerized SSH case naturally hold: whereas to calculate the standard entropies of an interval we merely have to count the cut bonds giving rise to the notion that the defect result is half of the topological result, this cannot be the case  for the SRE since $S_{n}^{\text{def}}\!\left(J^z_A=0\right)=0$ while $S_{n}^{\text{top}}\!\left(J^z_A=0\right)=\log 2$.  

Regarding the make up of the total entropy in terms of the configurational and fluctuational parts, we can write
\begin{equation}
S^{\text{triv}}_f=S^{\text{triv}}_c=0
\end{equation}
for the trivial phase, 
\begin{equation}
\begin{split}
S^{\text{top}}_{f}&=\frac{3}{2}\log2\\
S^{\text{top}}_{c}&=\frac{1}{2}\log2
\end{split}
\end{equation}
for the topological phase, and finally 
\begin{equation}
\begin{split}
S^{\text{def}}_{f}&=\log2\\
S^{\text{def}}_{c}&=0\,.
\end{split}
\end{equation}
The conclusions drawn at the discussion of the fully dimerized case of the SSH chain remain valid again as well.

\subsection{The two $J^z_\text{tot}=0$ ground states and their hybridization}

We can also study the two $J^z_\text{tot}=0$ ground states, i.e., $\mid \Updownarrow \rangle_+$ and $\mid \Updownarrow \rangle_-$, or their linear combinations which we write as 
\begin{equation}
\mid \Updownarrow \rangle_{p}=\sqrt{1-p}\mid \Updownarrow \rangle_+ +\sqrt{p}\mid \Updownarrow \rangle_-\,.
\end{equation}
Similarly to the SSH case, the only scenario to work out is when the interval contains the defect. 
If the interval does not contain the defect, the corresponding formulas of the above subsection apply without changes, irrespective of whether we choose $\mid\Updownarrow \rangle_+$, $\mid\Updownarrow \rangle_-$ or $\mid \Updownarrow \rangle_{p}$. 
It is easy to show that for the case of the defect the reduced density matrix can be essentially written in terms of two spin-1/2 degrees of freedom in the following way
\begin{equation}
\begin{split} 
& \rho=\frac{1}{4}\mathbb{I}-\frac{\sqrt{\left(1-p\right) p}}{2}
\left(\begin{array}{cccc}
 1 & 0 & 0 & 0\\
0 &  1 & 0 & 0\\
0 & 0 & - 1 & 0\\
0 & 0 & 0 & - 1\\
\end{array}\right)\,,
\end{split}
\end{equation}
where $\mathbb{I}$ is the $4\times4$ identity matrix and when the basis is made up by $\mid\uparrow\uparrow\rangle$, $\mid\downarrow\uparrow\rangle$, $\mid\uparrow\downarrow\rangle$ and $\mid\downarrow\downarrow\rangle$, i.e., the conventional basis of two spin-1/2 degrees of freedom.
This means that, introducing 
\begin{equation}
\eta=2\sqrt{p(1-p)}\
\end{equation}
for simplicity,
the spin resolved entropies can be expressed as
\begin{equation}
S^{\left(\eta\right)}_{n}\!\left(J_A^z=0\right)=\frac{1}{1-n}\log\left(\left(\frac{1+\eta}{2}\right)^{\!n}+\left(\frac{1-\eta}{2}\right)^{\!n}\right),
\end{equation}
and they vanish for $J^z_A\neq 0$. 
In particular the von Neumann SRE reads
\begin{equation}
S^{\left(\eta\right)}\!\left(J_A^z=0\right)=-\frac{1+\eta}{2}\log\!\left(\frac{1+\eta}{2}\right)-\frac{1-\eta}{2}\log\!\left(\frac{1-\eta}{2}\right),
\end{equation}
and it also vanishes for $J^z_A\neq 0$.

The above formulas are essentially the same as those for the SSH chain, just for the AKLT model we used the symmetric and anti-symmetric eigenstates to construct $\mid \Updownarrow \rangle_{p}$, whereas for the SSH model the interpolating state was defined via localized zero modes rather than their symmetric and anti-symmetric combinations. In other words, the role of the parameter $p$ is slightly different in the two cases and the $p$ parameter of the SSH case corresponds to the $\eta$ for the AKLT chain.

Finally, the total entropy is
\begin{equation}
S^{\left(\eta\right)}_{n}=\log2+\frac{1}{1-n}\log\left(\left(\frac{1+\eta}{2}\right)^{\!n}+\left(\frac{1-\eta}{2}\right)^{\!n}\right),
\end{equation}
and in particular
\begin{equation}
\begin{split}
S^{\left(\eta\right)}=&\log2-\frac{1+\eta}{2}\log\!\left(\frac{1+\eta}{2}\right)\\
&-\frac{1-\eta}{2}\log\!\left(\frac{1-\eta}{2}\right).
\end{split}
\end{equation}
We can observe that the change in the entropies w.r.t.~the non-hybridized ground states $\mid \Updownarrow \rangle_+$ and $\mid \Updownarrow \rangle_-$ is analogous to what was found for the SSH chase, that is,
\begin{equation}
\begin{split}
\Delta S^{\left(\eta\right)}&=\Delta S^{\left(\eta\right)}(J_A^z=0)\\
&=-\frac{1+\eta}{2}\log\!\left(\frac{1+\eta}{2}\right)-\frac{1-\eta}{2}\log\!\left(\frac{1-\eta}{2}\right),
\end{split}
\end{equation}
and
\begin{equation}
\begin{split}
\Delta S^{\left(\eta\right)}_n&=\Delta S^{\left(\eta\right)}_n(J_A^z=0)\\
\vspace{0.5cm}
&=\frac{1}{1-n}\log\left(\left(\frac{1+\eta}{2}\right)^{\!n}+\left(\frac{1-\eta}{2}\right)^{\!n}\right).
\end{split}
\end{equation}

\end{document}